\documentclass[twocolumn,showpacs,preprintnumbers,amsmath,amssymb,prb]{revtex4}

\usepackage{graphicx}
\usepackage{dcolumn}
\usepackage{epsf}

\begin{document}

\title{Enhancement of the upper critical field by nonmagnetic impurities in dirty two-gap superconductors }
\author{A. Gurevich}
\affiliation{Applied Superconductivity Center, University of Wisconsin, Madison, Wisconsin 53706}

\date{\today}

\begin{abstract}
Quasiclassic Uzadel equations for two-band superconductors in the dirty limit with the account of both intraband 
and interband scattering by nonmagnetic impurities are derived for any anisotropic Fermi surface. From these 
equations the Ginzburg-Landau equations, and the critical temperature $T_c$ are obtained. An equation for 
the upper critical field, which determines both the temperature dependence of  $H_{c2}(T)$ and the orientational 
dependence of $H_{c2}(\theta)$ as a function of the angle $\theta$ between ${\bf H}$ and the c-axis is obtained. 
It is shown that the shape of the $H_{c2}(T)$ curve essentially depends on the ratio of the intraband electron diffusivities 
$D_1$ and $D_1$, and can be very different from the standard one-gap dirty limit theory. In particular, the value 
$H_{c2}(0)$ can considerably exceed $0.7T_cdH_{c2}/dT_c$, which can have important consequences for applications 
of $MgB_2$. A scaling relation is proposed which enables one to obtain the angular dependence of $H_{c2}(\theta)$ from 
the equation for $H_{c2}$ at ${\bf H}\| c$. It is shown that, depending on the relation between $D_1$ and $D_2$, 
the ratio of the upper critical field $H_{c2}^\|/H_{c2}^\perp$ for ${\bf H}\| ab$ and ${\bf H}\perp ab$ 
can both increase and decrease as the temperature decreases. Implications of the obtained results for $MgB_2$ 
are discussed.

\end{abstract}
]

\maketitle

\section{Introduction}
Recent discovery of $MgB_2$ has renewed interest in superconductors with multicomponent order 
parameters. In particular, extensive ab-initio calculations \cite{tg1,tg2}, along with ample experimental evidences 
based on STM \cite{tg3,tg4}, point contact \cite{tg5}, 
and Raman \cite{tg6} spectroscopy, heat capacity \cite{tg7} and rf response \cite{tg8} measurements strongly indicate 
that $MgB_2$ has two distinct s-wave superconducting gaps $\Delta_\sigma (0)\approx 7.2$mV and $\Delta_\pi (0)\approx 2.3$mV.
These gaps reside on different disconnected sheets of the Fermi surface (FS), which comprises nearly cylindrical 2D parts formed by 
in-plane $\sigma$ antibonding $p_{xy}$ orbitals of B, and a more isotropic 3D tubular network formed by out-of-plane $\pi$ bonding and 
antibonding $p_z$ orbitals of B.  So far all attempts to increase the critical temperature $T_c$ of $MgB_2$ by doping have been 
unsuccessful \cite{rev}. On the other hand, the significant potential of $MgB_2$ for applications\cite{appl} is limited by comparatively low upper 
critical fields $H_{c2}^{\perp}(0)\simeq 3-5T$ and  $H_{c2}^{||}(0)\simeq 15-19T$ of hexagonal $MgB_2$ single crystals \cite{a1,a2,a3,a4,a5,a6,a7,a8,a9,a10,a11}, where the indices $\perp$ and $||$ correspond to  the magnetic field {\bf H} perpendicular 
and parallel to the ab plane, respectively. As far as $H_{c2}$ is concerned, it can be increased by adding nonmagnetic 
impurities \cite{whh,a15}, which are especially effective in the dirty limit, $2\pi k_BT_c < \hbar/\tau$, where $\tau$ is the elastic 
scattering time, $\hbar$ is the Planck constant, and $k_B$ is the Boltzmann constant. In this case, there is a simple universal 
relation between the zero-temperature value $H_{c2}(0)$, the slope $H_{c2}'=dH_{c2}/dT$ at $T_c$, and the normal state 
residual resistivity $\rho_n$ \cite{whh}:
	\begin{equation}		
	H_{c2}(0)=0.69T_cH_{c2}', \qquad H_{c2}'=\frac{4ek_B}{\pi\hbar}N_F\rho_n,
	\label{whh}
	\end{equation} 
where $N_F$ is the density of states at the FS and $-e$ is the electron charge. The fact that adding  
nonmagnetic impurities increases $H_{c2}$ has been used to significantly improve $H_{c2}$ of $NbTi$ 
and $Nb_3Sn$ compounds \cite{a15}. The same approach has been also taken to improve the 
high-field performance of $MgB_2$ in which the additional scattering was introduced by proton \cite{perk} and neutron \cite{buket} 
irradiation, and atomic substitutions on both B and Mg sites in thin films \cite{sat}, bulk samples \cite{rev} and wires \cite{jc1,jc2}. 
For instance, in highly resistive c-axis oriented $MgB_2$ films \cite{sat}, $\rho_n$ was increased from the nominal value 
$\sim 1 \mu\Omega$cm \cite{fin} to more than $700 \mu\Omega$cm, which in turn increased the slopes $H_{c2}'$ to 
rather high values $H_{c2\perp}'\approx 1$T/K and $H_{c2\|}'\approx 1.8$T/K, while reducing $T_c$ down to 
$\approx 30$K. Based on these numbers, the extrapolation (\ref{whh}) would give the upper critical field, 
$H_{c2}^\perp (0)\approx 20$T, which is still below $H_{c2}(0)\approx 30$T of $Nb_3Sn$\, \cite{a15}. However, 
Eq. (\ref{whh}) is a result of the one-gap theory \cite{whh} which does not take into account the multiple scattering channels 
in two-gap superconductors. In this paper I show that $H_{c2}(0)$ in two gap superconductors can be significantly higher than 
what follows from Eq. (\ref{whh}), thus the upper critical fields of $MgB_2$ alloys can in fact exceed $H_{c2}(0)$ of $Nb_3Sn$ even 
for the values $H_{c2\perp}'\approx 1$T/K which have already been achieved\cite{sat,gur}.
   	
The theory of the two-band superconductivity has been developed long ago \cite{suhl}, and more recently generalized to include 
the effects of impurities\cite{imp1,imp2,imp3,lanis1,lanis2,goluba} and strong electron-phonon coupling 
\cite{tg2,lanis2,goluba,shulga}. The Fermi surface of $MgB_2$ provides three different impurity scattering channels: intraband scattering within 
each $\sigma$ and $\pi$ FS sheets and interband scattering between them. Strong impurity scattering in the dirty limit causes intraband electron diffusion over the respective FS sheets, which reduces the intrinsic anisotropy of $\Delta_\sigma$ and $\Delta_\pi$, but does not affect $T_c$, in accordance with the Anderson theorem. By contrast, the pairbreaking interband scattering reduces $T_c$ \cite{imp1,imp2}, but its effect in $MgB_2$ seems to be not very pronounced due to orthogonality of $\sigma$ and $\pi$ orbitals\cite{imp3}. It is the multiple scattering channels, which 
provide the essential flexibility to increase the $H_{c2}$ of $MgB_2$ to a much greater extent than in one-gap superconductors not only by the usual increase of $\rho_n$, but also by optimizing the relative weight of the $\sigma$ and $\pi$ scattering rates. The latter could be done by 
selective atomic substitution on B sites (with C, O, etc. ) which mostly affect the in-plane $\sigma$ scattering, and Mg site 
(Mg vacancies, Al, etc) \cite{rev}, which mostly affects out-of-plane $\pi$ scattering \cite{imp3}. 

In this work I use a weak coupling multiband BCS model \cite{suhl} to develop a theory of magnetic properties 
of two-gap superconductors in the dirty limit, taking $MgB_2$ as an example. This approach 
generalizes the well-known theory developed for dirty one-gap superconductors\cite{whh}. The paper is organized as follows. In section II the 
quasiclassic Uzadel equations are obtained for an anisotropic two-gap superconductor with the account of both interband and 
intraband scattering by nonmagnetic impurities. In section III these equations are used to derive the Ginzburg-Landau 
equations and the critical temperature $T_c$ in the dirty limit. In section IV the linearized Uzadel equations are solved 
exactly to obtain a general equation for the upper critical field parallel to the c-axis. It is shown that the temperature dependence of 
$H_{c2}(T)$ can be very different from the one-band theory \cite{whh}, and $H_{c2}(0)$ can be considerably 
higher than what follows from Eq. (\ref{whh}).  In section V I calculate the angular dependence of $H_{c2}(\theta)$  
and propose a scaling rule, which shows that the anisotropy ratio $H_{c2}^\|/H_{c2}^\perp$ can both 
increase and decrease with temperature, depending on the relation between intraband diffusivities. 
In section VI the results of this work are used to address the observed temperature and orientational dependences 
of $H_{c2}$ of $MgB_2$. 

\section{Uzadel equations}

A powerful tool for investigating inhomogeneous states of superconductors is the quasiclassic 
Eilenberger equations \cite{eilen,book} for the Green functions $f({\bf k, r},\omega)$, $f^+({\bf k, r},\omega)$, $g({\bf k, r},\omega)$ 
and the order parameters $\Delta({\bf k, r})$ which depend on the coordinates ${\bf r}$, the Matsubara frequency 
$\omega = \pi T(2n+1)$, and the wave vector ${\bf k}$ on the FS (hereafter the units with $k_B=\hbar=1$ are used).  
The essential dependence of $f({\bf k, r},\omega)$, $g({\bf k, r},\omega)$ and $\Delta({\bf k, r})$ on the 
direction of ${\bf k}$, makes them sensitive to the shape of the FS, which greatly complicates solving the nonlinear 
Eilenberger equations. Thus, the Eilenberger equations are not so useful in the case of complicated FS, 
such as the tubular network of disconnected FS sheets of $MgB_2$ for which the numerical solutions of the Eliashberg 
equation have shown the significance of the FS anisotropy \cite{tg2}. However the situation simplifies 
in the dirty limit for which the impurity scattering within each FS sheet averages out the angular dependences of 
$f({\bf k, r},\omega)$, $f^+({\bf k, r},\omega)$, $g({\bf k, r},\omega)$, making them independent of ${\bf k}$. In this case the 
Eilenberger equations reduce to much simpler Uzadel equations \cite{uzadel} in which all 
microscopic details are hidden in the electronic diffusivity tensors $D_m^{\alpha\beta}$ for each m-th FS sheet 
and the interband scattering rates $\gamma_{mm'}$ which reflect the underlying FS symmetry \cite{allen}. 

In this paper I consider a dirty two-gap anisotropic superconductor 
in the simplest case of two disconnected sheets 1 and 2 of the FS. This model corresponds to the $\sigma$ and $\pi$ bands in 
$MgB_2$ \cite{tg1,tg2}, for which the superconducting gaps take the constant values $\Delta_1$ and $\Delta_2$ on the sheets 1 
and 2, respectively. The Uzadel equations derived in Appendix A have the form 
	\begin{eqnarray}
	2\omega f_1-D_1^{\alpha\beta}[g_1\Pi_\alpha\Pi_\beta f_1-f_1\nabla_\alpha\nabla_\beta g_1]=
	\nonumber \\
	2\Delta_1g_1+\gamma_{12}(g_1f_2-g_2f_1)\qquad
	\label{uz1} \\
	2\omega f_2-D_2^{\alpha\beta}[g_2\Pi_\alpha\Pi_\beta f_2-f_2\nabla_\alpha\nabla_\beta g_2]=
	\nonumber \\
	2\Delta_2g_2+\gamma_{21}(g_2f_1-g_1f_2),\qquad
	\label{uz2}
	\end{eqnarray}
Eqs. (\ref{uz1}) and (\ref{uz2}) are supplemented by the self-consistency equations for the order parameters 
$\Delta_m=|\Delta_m|\exp(i\varphi_m)$: 
	\begin{equation}
	\Delta_m=2\pi T\sum_{\omega > 0}^{\omega_D}\sum_m\lambda_{mm'}f_{m'}({\bf r}, \omega),
	\label{d}
	\end{equation} 
normalization condition
	\begin{equation}
	|f_m|^2+g_m^2=1,
	\label{norm}
	\end{equation}
and the expression for the supercurrent density
	\begin{equation}
	J^\alpha=4\pi i eT Im\sum_\omega\sum_m N_mD_m^{\alpha\beta}f_m^+\Pi_\beta f_m.
	\label{j}
	\end{equation}
Here the band index m runs from 1 and 2, the functions $f_m({\bf r},\omega)$ and $g_m({\bf r},\omega)$ in the 
m-th band depend on ${\bf r}$ and $\omega$ but not on ${\bf k}$, $D_m^{\alpha\beta}$ are the intraband diffusivity tensors 
due to nonmagnetic impurity scattering, $\gamma_{mm'}$ are the interband scattering rates, 
$N_m$ is the partial density of states, ${\bf \Pi}=\nabla + 2\pi i {\bf A}/\phi_0$, ${\bf A}$ is the vector potential, 
$\phi_0$ is the flux quantum, and the summation over the Greek Cartesian indices is implied. 

Eqs. (\ref{d}) contains the matrix of the BCS superconducting coupling constants 
$\lambda_{mm'}=\lambda_{mm'}^{(ep)}-\mu_{mm'}$, where  $\lambda_{mm'}^{(ep)}$ are electron-phonon constants, and 
$\mu_{mm'}$ is the matrix of the Coulomb pseudopotential. Here the diagonal terms $\lambda_{11}$ and $\lambda_{22}$ quantify the 
intraband superconducting coupling, and off-diagonal terms $\lambda_{12}$ and $\lambda_{21}$ describe the interband 
coupling.   The eigenvalues of $\lambda_{mm'}$ are assumed positive, the band 1 having the highest coupling constant 
$\lambda_{11}$. The indices 1 and 2 thus correspond to the 
$\sigma$ and $\pi$ bands of $MgB_2$ for which {\it ab-initio} calculations yield 
$\lambda_{\sigma\sigma}\approx 0.81$, $\lambda_{\pi\pi}\approx 0.285$, $\lambda_{\sigma\pi}\approx 0.119$, 
and $\lambda_{\pi\sigma}\approx 0.09$\, \cite{goluba}. The mixed components $\lambda_{12}$ and $\lambda_{21}$ 
satisfy the symmetry relation \cite{suhl} (see also Eq. (\ref{a22})):
	\begin{equation}
	N_1\lambda_{12}=N_2\lambda_{21}
	\label{inter}
	\end{equation} 
where $N_1$ and $N_2$ are partial densities of states in the bands 1 and 2 ($N_\pi\approx 1.3N_\sigma$ for $MgB_2$). 
A similar approach based on the Uzadel equations was recently proposed 
to describe vortices in two-gap superconductors \cite{koshgol}. 

Formulas for the interband scattering rates $\gamma_{mm'}$ and the intraband diffusivity tensors $D_m^{\alpha\beta}$ 
expressed in terms of microscopic material parameters are given in Appendix A. They will be used in the next sections as 
input parameters, which can be either calculated from first principles or extracted from 
the observed temperature dependences of $H_{c2}(T)$ and resistivity $\rho$, as discussed below. Here we just emphasize 
the fact that, although both $D_1^{\alpha\beta}$ and $D_2^{\alpha\beta}$ reflect the underlying symmetry of the FS, 
the features of atomic orbitals which form the bands 1 and 2 can manifest themselves in very different properties of  
$D_1^{\alpha\beta}$ and $D_2^{\alpha\beta}$. For instance, the principal value $D_\sigma^{(c)}$ along the c-axis in $MgB_2$ 
is much smaller than two principal values $D_\sigma^{(a)}$ and $D_\sigma^{(b)}$ in the ab plane, because of the nearly 
2D nature of the $\sigma$-band formed by in-plane bonding and antibonding $p_{xy}$ orbitals of B, as was  
shown by STM \cite{tg3}. By contrast, the difference in the principal values  
$D_\pi^{(c)}$, $D_\pi^{(a)}$ and $D_\pi^{(b)}$ is less pronounced for the 3D $\pi$-band formed by the out-of-plane $p_z$ 
orbitals of B. The resulting relation $D_\sigma^{(c)}/D_\sigma^{(a)}\ll D_\pi^{(c)}/D_\pi^{(a)}$ can manifest itself in anomalous 
temperature dependence of the anisotropy of $H_{c2}$, as shown in the next sections.       

\section{Critical temperature and Ginzburg-Landau equations}

The necessity to satisfy the self-consistency condition (\ref{d}) complicates solving the nonlinear Eqs. (\ref{uz1})-(\ref{uz2}) 
for inhomogeneous $\Delta_m({\bf r})$. The situation simplifies near $T_c$ where  
Eqs. (\ref{uz1})-(\ref{d}) reduce to the Ginzburg-Landau (GL) equations for the order parameters $\Delta_m$. 
We derive the GL equations neglecting the interband scattering terms $\gamma_{12}\ll \pi T_c$ which are usually small in 
$MgB_2$ due to orthogonality of the $\sigma$ and $\pi$ orbitals \cite{imp3}. Expanding the solution of Eqs. (\ref{uz1}) and (\ref{uz2}) 
in powers of $\Delta_m$ and its spatial derivatives, and using $g_m\simeq 1-|f_m|^2/2$, we obtain 
	\begin{equation}
	f_m=\frac{\Delta_m}{\omega}+\frac{D_m^{\alpha\beta}}{2\omega^2}{\Pi_\alpha\Pi_\beta \Delta_m}-\frac{\Delta_m|\Delta_m|^2}{2\omega^3}
	\label{expans}
	\end{equation}  
Inserting Eqs. (\ref{expans}) into Eq. (\ref{d}) and summing up over $\omega$, results in the following equations
	\begin{eqnarray}
	\Delta_1=\lambda_{11}R_1+\lambda_{12}R_2, \qquad\qquad\qquad
	\label{r1} \\
	\Delta_2=\lambda_{21}R_1+\lambda_{22}R_2, \qquad\qquad\qquad
	\label{r2} \\
	R_m=\Delta_m l+\frac{\pi D_m^{\alpha\beta}}{8T}
	\Pi_\alpha\Pi_\beta\Delta_m-\frac{7\zeta(3)}{8\pi^2T^2}|\Delta_m|^2\Delta_m,
	\label{rm}
	\end{eqnarray}
where $R_m=\sum_{\omega}f_m$, $l=\ln(2\gamma\omega_D/\pi T)$, $\ln\gamma = -0.577$ is the Euler constant, 
and $\zeta(3)=1.202$. The supercurrent density at $T-T_c\ll T_c$ is obtained by inserting the 
first term in the right hand side of Eq. (\ref{expan}) into Eq. (\ref{j}) and summing up over $\omega$:
	\begin{equation}
	J^\alpha=-\frac{\pi e}{4T}Im\sum_mN_mD_m^{\alpha\beta}\Delta_m^*\Pi_\beta\Delta_m,
	\label{jgl}
	\end{equation}
where the asterisk means complex conjugation. As follows from Eq. (\ref{jgl}), the current density 
near $T_c$ is a sum of independent intraband contributions. 

Equilibrium equations (\ref{r1})-(\ref{r2}) for the order parameters  $\Delta_m$ are not 
yet the GL equations which are obtained by varying the free energy functional, $\delta F/\delta\Delta^*=0$. The reason 
is that the off-diagonal terms with respect to the band index m in Eqs. (\ref{r1})-(\ref{r2})
violate the necessary symmetry conditions which the  variational equations $\delta F/\delta\Delta^*=0$ must satisfy. 
To find the proper linear combinations of Eqs. (\ref{r1}) and (\ref{r2}) which satisfy all necessary symmetry conditions, 
we first solve Eqs. (\ref{r1})-(\ref{rm}) for the quantities $N_1R_1=(\lambda_{22}\Delta_1-\lambda_{12}\Delta_2)N_1/w$ and 
$N_2R_2=(\lambda_{11}\Delta_2-\lambda_{21}\Delta_1)N_2/w$, where $w=\lambda_{11}\lambda_{22}-\lambda_{12}\lambda_{21}$ 
is the determinant of the matrix $\lambda_{mm'}$. Expressing then $R_m$ via $\Delta_m$ using Eq. (\ref{rm}), we obtain 
the GL equations for an anisotropic two-band superconductor:
	\begin{eqnarray}
	-a_1\Delta_1+b_1\Delta_1|\Delta_1|^2 + c_1^{\alpha\beta}\Pi_\alpha\Pi_\beta\Delta_1+g\Delta_2=0,
	\label{gl1} \\
	-a_2\Delta_2+b_2\Delta_2|\Delta_2|^2 + c_2^{\alpha\beta}\Pi_\alpha\Pi_\beta\Delta_2+g\Delta_2=0.
	\label{gl2}
	\end{eqnarray}
Here the GL expansion coefficients are given by
	\begin{eqnarray}
	a_1=\left(\ln\frac{2\gamma\omega_D}{\pi T}-\frac{\lambda_{22}}{w}\right)N_1, \quad b_1=\frac{7\zeta(3)N_1}{8\pi^2T^2},
	\label{coef1} \\
 	a_2=\left(\ln\frac{2\gamma\omega_D}{\pi T}-\frac{\lambda_{11}}{w}\right)N_2, \quad b_2=\frac{7\zeta(3)N_2}{8\pi^2T^2}, 
	\label{coef2}\\
	c_1^{\alpha\beta}=\pi N_1D_1^{\alpha\beta}/8T,\qquad c_2^{\alpha\beta}=\pi N_2D_2^{\alpha\beta}/8T,
	\label{coef3} \\
	g=\lambda_{21}N_2/w=\lambda_{12}N_1/w.\qquad\qquad
	\label{coef4}
	\end{eqnarray}
The equality in Eq. (\ref{coef4}) results from Eq. (\ref{inter}) for the interband coupling constants $\lambda_{12}=V_iN_2$, and 
$\lambda_{21}=V_iN_1$. The GL equations (\ref{gl1}) and (\ref{gl2}) can now be obtained by varying the free energy 
$F=\int[F_1+F_2+F_i]dV$, which contains the usual GL intraband contributions 
	\begin{equation}
	F_m=-\frac{a_m}{2}|\Delta_m|^2 +\frac{b_m}{4}|\Delta_m|^4+\frac{c_m^{\alpha\beta}}{2}\Pi_\alpha\Delta_m\Pi_\beta^*\Delta^*_m,\qquad\qquad
	\label{fm}
	\end{equation}
and the interband interaction term
	\begin{equation}
	F_i=\frac{g}{4}(\Delta_1\Delta_2^*+\Delta_2\Delta_1^*).
	\label{fi}
	\end{equation}  	
Notice that Eq. (\ref{jgl}) for the current density obtained from the Uzadel equations coincides  
with the general expression ${\bf J}=\delta F/\delta{\bf A}$, where $F$ is defined by Eqs. (\ref{fm}) and (\ref{fi}). 
Static GL equations (\ref{gl1})-(\ref{fi}) and their time-dependent generalization \cite{gurvin} have been suggested 
phenomenologically to describe various distributions of the order parameter and vortex properties in two-gap 
superconductors\cite{tanaka,gurvin}. 

Eqs. (\ref{gl1}) and (\ref{gl2}) give the following equation $a_1a_2-g^2=0$ for the critical temperature $T_c$, 
which results in the quadratic equation for $l_c=\ln(2\gamma\omega_D/\pi T_c)$: 
	\begin{equation}
	1-\lambda_+l_c+wl_c^2=0
	\label{quadeq}
	\end{equation}
where $\lambda_{\pm}=\lambda_{11}\pm\lambda_{22}$, and $w=\lambda_{11}\lambda_{22}-\lambda_{12}\lambda_{21}$. 
Solution of Eq. (\ref{quadeq}) reproduces the well-known result by Suhl, Matthias and Walker \cite{suhl}: 
	\begin{equation}
	T_{c0}=1.14\omega_D\exp[-(\lambda_+-\lambda_0)/2w], 
	\label{suhl}
	\end{equation}
where $\lambda_0=(\lambda_-^2+4\lambda_{12}\lambda_{21})^{1/2}$. 
As follows from Eq. (\ref{suhl}), the interband coupling 
always increases $T_c$ as compared to noninteracting bands $(\lambda_{12}=\lambda_{21}=0)$, while the 
intraband impurity scattering does not affect $T_c$, in accordance with the Anderson theorem \cite{book}.

\section{Upper critical field for ${\bf H}\| c$}

Now we turn to the calculation of the upper critical field $H_{c2}$ applied along the c-axis of a 
hexagonal crystal. Here $H_{c2}$ is the maximum eigenvalue of the linearized Eqs. (\ref{uz1}) and 
(\ref{uz2}) for $f\ll 1$ and $g\to 1$. In Appendix B, these equations are solved exactly 
and an equation for $H_{c2}$ is obtained for arbitrary relation between $D_1$, $D_2$ and 
$\gamma_{mm'}$. This general equation for $H_{c2}$ is very cumbersome, so to make the 
essential effects of two-gap superconductivity more transparent, I consider here a much simpler 
case of negligible $\gamma_{mm'}$ for which the nonmagnetic impurity scattering does not affect 
$T_c$. If $\gamma_{12}=\gamma_{21}=0$, the linearized Eqs. (\ref{uz1}) and (\ref{uz2}) take the form
	\begin{eqnarray}
	2\omega f_1-D_1^{\alpha\beta}\Pi_\alpha\Pi_\beta f_1=2\Delta_1, 
	\label{uzl1} \\
	2\omega f_2-D_2^{\alpha\beta}\Pi_\alpha\Pi_\beta f_2=2\Delta_2.
	\label{uzl2} 
	\end{eqnarray}
For ${\bf H}|| c$, the relevant solutions of  
Eqs. (\ref{uzl1}) and (\ref{uzl2}) depends only on isotropic in-plane diffusivities, 
$D_m^{\alpha\beta}=D_m\delta_{\alpha\beta}$. In the gauge $A_y=Hx$ 
the solutions of Eqs. (\ref{d}), (\ref{uzl1}), and (\ref{uzl2}) are 
	\begin{eqnarray}
	f_m(x,\omega)=\Delta_m(x)/(\omega+\pi H D_m/\phi_0),  
	\label{fom} \\
 	\Delta_m(x)=\tilde{\Delta}_m\exp(-\pi Hx^2/\phi_0).
	\label{dom}
	\end{eqnarray} 
Inserting Eqs. (\ref{fom}) and (\ref{dom}) into the gap equation (\ref{d}) yields
two linear equations for $\tilde{\Delta}_1$ and  $\tilde{\Delta}_2$ in which the  
summation over $\omega$ is performed using the identity 
	\begin{eqnarray}
	2\pi T\sum_{\omega>0}^{\omega_D}\frac{1}{\omega+\Omega}=\ln\frac{2\gamma\omega_D}{\pi T}-U\left(\frac{\Omega}{2\pi T}\right), 
	\label{ident} \\
	U(x)=\psi (1/2+x)+\psi (1/2), \qquad
	\label{udef}
	\end{eqnarray}	
where $\psi(x)$ is the di-gamma function. The equations for $\tilde{\Delta}_1$ and $\tilde{\Delta}_2$ 
become 
	\begin{eqnarray}
	\tilde{\Delta}_1=\lambda_{11}[l-U(h)]\tilde{\Delta}_1+\lambda_{12}[l-U(\eta h)]\tilde{\Delta}_2, 
	\label{delt1} \\
	\tilde{\Delta}_2=\lambda_{22}[l-U(\eta h)]\tilde{\Delta}_2+\lambda_{21}[l-U(h)]\tilde{\Delta}_1,
	\label{delt2} 
	\end{eqnarray}
where $l=\ln(2\gamma\omega_D/\pi T)$, $h=H_{c2} D_1/2\phi_0 T$, and $\eta=D_2/D_1$. 
The solvability condition of Eqs. (\ref{delt1}) and (\ref{delt2}) gives an equation for $H_{c2}$, in which 
it is convenient to express $\omega_D$ via $T_c$ using Eqs. (\ref{quadeq}) and (\ref{suhl}). As a result, 
the equation for $H_{c2}$ takes the form
	\begin{eqnarray}
	2a_0[\ln t+ U(h)][\ln t+ U(\eta h)]+
	\nonumber \\
	a_2[\ln t+U(\eta h)]+a_1[\ln t+U(h)]=0.\quad
	\label{hc2}
	\end{eqnarray}
where $a_1 = 1 + \lambda_- /\lambda_0$, $a_2=1-\lambda_- /\lambda_0$, and $a_0=w/\lambda_0$.
For  equal diffusivities, $\eta = 1$, Eq. (\ref{hc2}) reduces to the equation  
$\ln t + U(h)=0$ for $H_{c2}$ in one-gap dirty superconductors \cite{whh}. A somewhat different form of 
Eq. (\ref{hc2}) was recently obtained from the Uzadel equations \cite{koshgol}.  

\begin{figure}          
\epsfxsize= 0.7\hsize
\centerline{
\vbox{
\epsffile{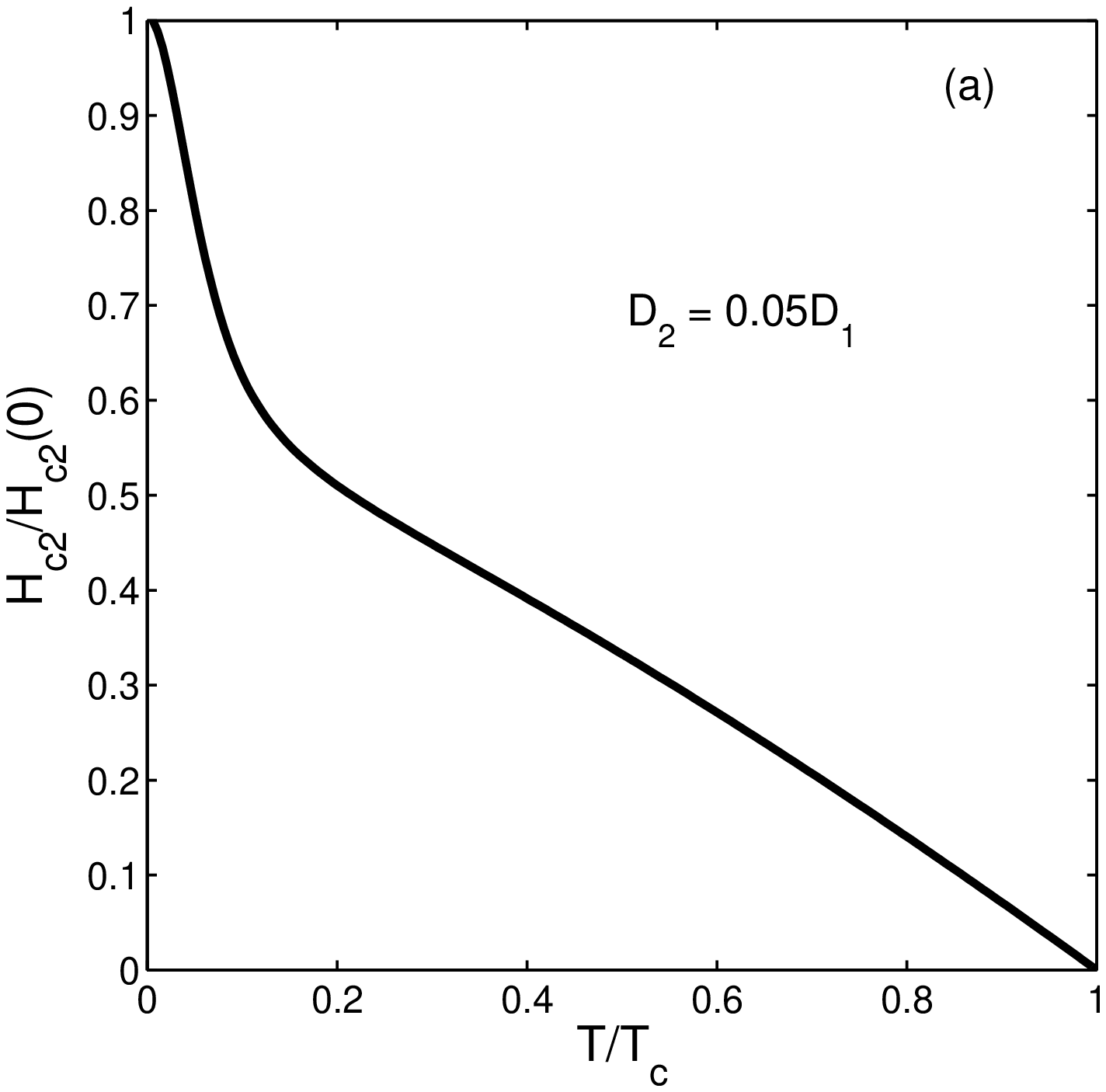}
}}
\epsfxsize= 0.7\hsize
\centerline{
\vbox{
\epsffile{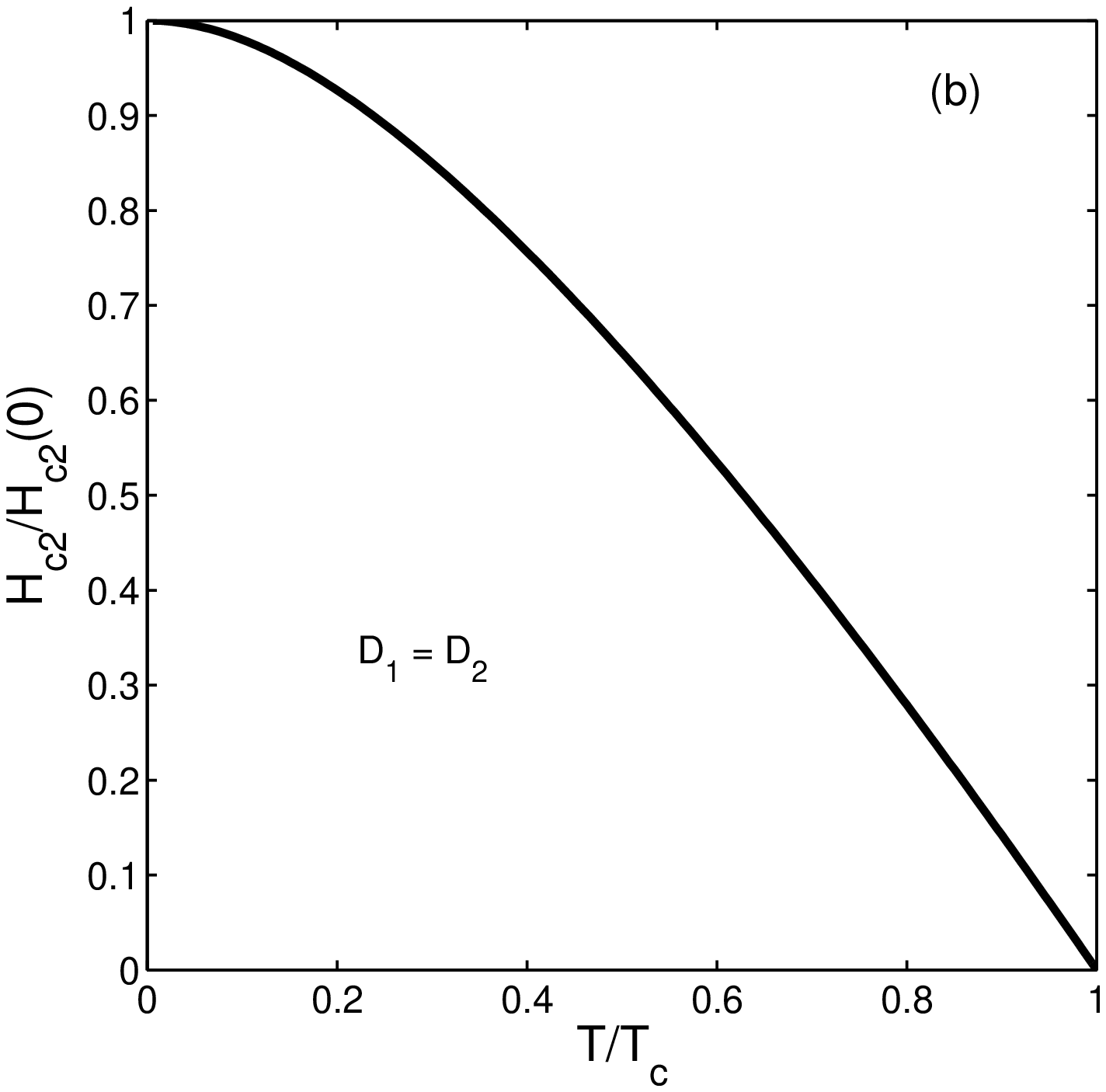}
}}
\epsfxsize= 0.7\hsize
\centerline{
\vbox{
\epsffile{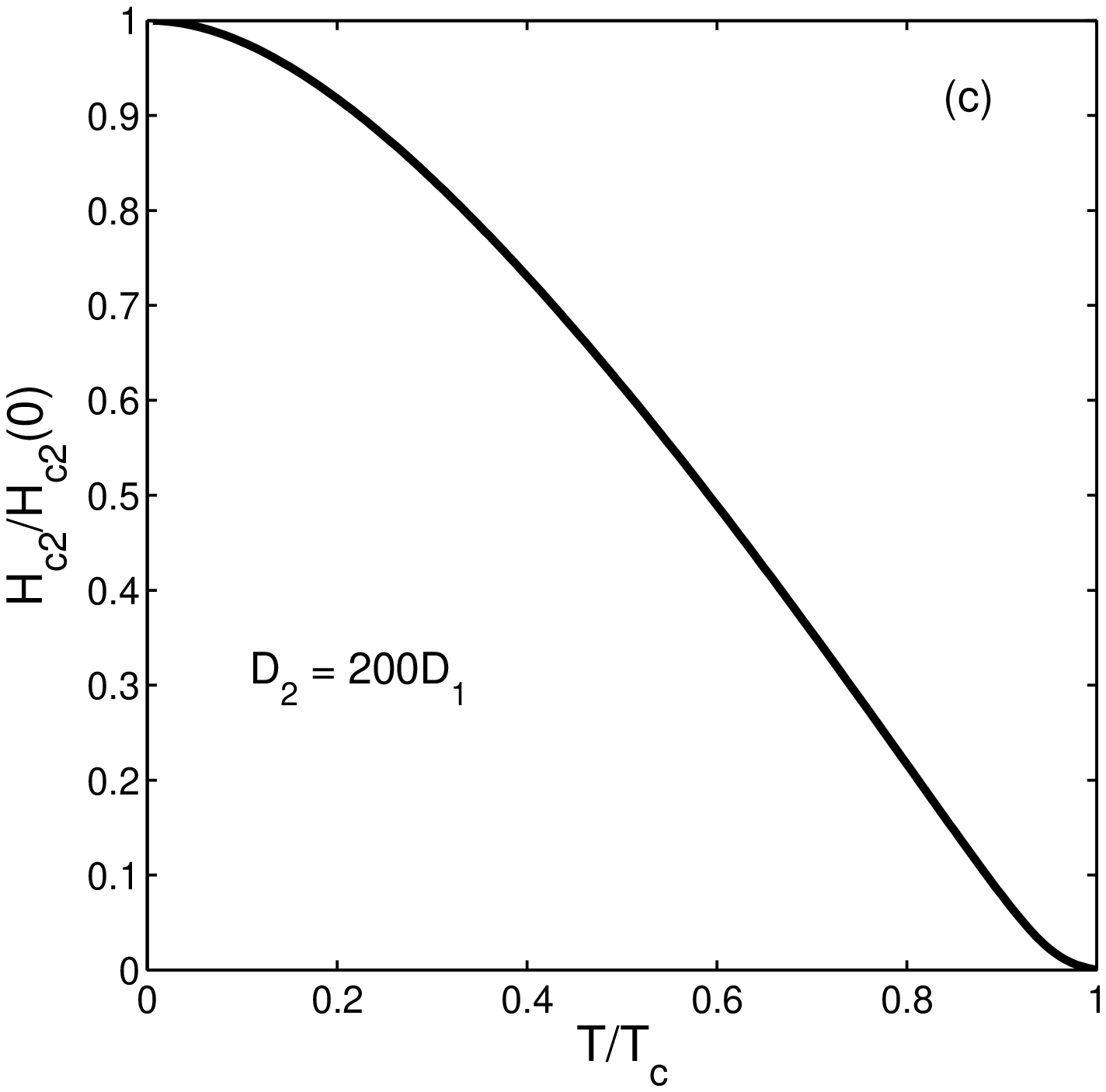}
}}

\vskip \baselineskip
\caption{Temperature dependences $H_{c2}(T)$ calculated from Eq. (\ref{hc2}) for different ratios of $D_2/D_1$ and 
the coupling constants $\lambda_{\sigma\sigma}\approx 0.81$, $\lambda_{\pi\pi}\approx 0.285$, $\lambda_{\sigma\pi}\approx 0.119$, $\lambda_{\pi\sigma}\approx 0.09$ calculated for $MgB_2$\, \cite{goluba}. 
}
\label{fig.1}
\end{figure}

Shown in Fig. 1 are $H_{c2}(T)$ curves calculated from Eq. (\ref{hc2}) for different diffusivity ratios $\eta=D_2/D_1$. 
One can clearly see the evolution of the temperature dependence of $H_{c2}(T)$ from the dirty-limit one-band  
behavior at $D_1=D_2$, to rather different $H_{c2}(T)$ curves which have portions with both upward and 
downward curvatures for either $D_1\ll D_2$ and $D_1\gg D_2$.  In the latter two cases, the zero-temperature value 
$H_{c0}(0)$ can be significantly higher than the one-gap extrapolation  (\ref{whh}). To show this,  
we obtain expressions for $H_{c2}(T)$ near $T_c$ and $T=0$. For $T\approx T_c$, Eq. (\ref{hc2}) can be 
expanded in small terms $\sim h$, using $U(h)\approx \pi^2h/2$, and then solving for $H_{c2}$: 
	\begin{equation}
	H_{c2}=\frac{8\phi_0(T_c-T)}{\pi^2(a_1D_1+a_2D_2)},
	\label{hc2c}
	\end{equation} 
For $D_1=D_2$, Eq. (\ref{hc2c}) yields the result of the one-band theory, 
$H_{c2}=4(T_c-T)\phi_0/\pi^2D$ \cite{whh}. Furthermore, if $\lambda_{12}=\lambda_{21}=0$, 
then $a_1=2$ and $a_2=0$, thus $H_{c2}(T)4(T_c-T)\phi_0/\pi^2D_1$ is determined by 
$D_1$ for the band with the highest coupling constant $\lambda_{11}$. However, if all 
constants $\lambda_{mm'}$ are finite, but $D_1$ and $D_2$ are very different, then 
$H_{c2}$ is determined by the maximum diffusivity, for example, by $D_2\ll D_1$ if the scattering 
in the "weak" band 2 dominates. 

The zero-temperature $H_{c2}(0)$ can be obtained using the asymptotic behavior of 
$U(x) \approx \ln (4\gamma h)$ for $h\to\infty$. Then all terms proportional to $\ln t$ cancel out, and Eq. (\ref{hc2}) 
reduces to a quadratic equation for $\ln H_{c2}$, whence   
	\begin{eqnarray}
	H_{c2}(0)=\frac{\phi_0T_c}{2\gamma\sqrt{ D_1 D_2}}\exp(\frac{g}{2}),\qquad\qquad
	\label{hc2o} \\
	g=\left( \frac{\lambda_0^2}{w^2}+\ln^2\frac{D_2}{D_1}+\frac{2\lambda_-}{w}\ln\frac{D_2}{D_1}\right)^{1/2}-\frac{\lambda_0}{w}.\quad
	\label{goo}
	\end{eqnarray}
For $D_1=D_2$, Eqs. (\ref{hc2o})-(\ref{goo}) again yield the result $H_{c2}(0)=\phi_0T_c/2\gamma D$ of the 
one-band dirty limit \cite{whh}.  However for $D_1\neq D_2$, Eqs. (\ref{hc2})-(\ref{goo}) predict a 
significant enhancement of $H_{c2}(0)$ as compared to $H_{c2}(0)=0.69H_{c2}'T_c$ for the symmetric case $D_1=D_2$. 
Indeed, in the limit of very different diffusivities, we obtain
	\begin{eqnarray}
	H_{c2}(0)=\frac{\phi_0T_c}{2\gamma D_2}e^{\frac{\lambda_- -\lambda_0}{2w}}, \qquad D_2\ll D_1e^{-\frac{\lambda_0}{w}},
	\label{o1} \\
	H_{c2}(0)=\frac{\phi_0T_c}{2\gamma D_1}e^{\frac{\lambda_- -\lambda_0}{2w}}, 
	\qquad D_1\ll D_2e^{-\frac{\lambda_0}{w}}.
	\label{o2}
	\end{eqnarray}
Unlike $H_{c2}(T)$ at $T\approx T_c$, the limiting value of $H_{c2}(0)$ is determined by the {\it minimum} 
diffusivity. It is the feature of a two-band superconductor, which causes both the upward 
curvature and the enhancement of $H_{c2}$ at low T in Fig. 1. In the limit $D_2\ll D_1$, the 
diffusivity $D_2$ does not affect $H_{c2}$ practically for all temperatures $T<T_c$, 
except a narrow region $T\ll T_c$, but for $D_2\to 0$, the upper critical field $H_{c2}(T)$ diverges at $T=0$. 
 
\section{Anisotropy of $H_{c2}$}.

\subsection{General equations}

To address the anisotropy of $H_{c2}$, 
observed in $MgB_2$, we consider a uniaxial crystal in a field ${\bf H}$ inclined by the angle $\theta$ with respect to the c-axis. 
Then the tensors $D_{\alpha\beta}$ in Eq. (\ref{uz1})-(\ref{uz2}) have two equal principal values $D_m^{(a)}$ in the ab 
plane and a different value $D_m^{c}$ along the c-axis. If the z-axis is taken along ${\bf H}$ and the a-axis coincides with the 
y-axis, $D_m^{\alpha\beta}$ has the following nonzero components 
	\begin{eqnarray}
	D_m^{yy}&=&D_m^{(a)}, 
	\label{dyy} \\ 
	D_m^{xx}&=&D_m^{(a)}\cos^2\theta+D_m^{(c)}\sin^2\theta, 
	\label{dxx}\\
	D_m^{zz}&=&D_m^{(c)}\cos^2\theta+D_m^{(a)}\sin^2\theta,
	\label{dzz} \\ 
	D_m^{xz}&=&D_m^{(zx)}=(D_m^{(c)}-D_m^{(a)})\sin\theta\cos\theta
	\label{dxz}
	\end{eqnarray}
In the gauge $A_y=Hx$, the Uzadel equations (\ref{uzl1}) and (\ref{uzl2}) take the form
	\begin{eqnarray}
	-D_m^{xx}\partial_{xx}f_m - D_m^{yy}(\partial_y+2\pi iHx/\phi_0)^2f_m \nonumber \\
	-D_m^{zz}\partial_{zz}f_m - 2D_m^{xz}\partial_{xz}f_m+2\omega f_m=2\Delta_m
	\label{anisuz}
	\end{eqnarray}
The solution of this equation has the form $f_m({\bf r})=\exp(ik_zz-ik_zxD_m^{xz}/D_m^{xx}+ik_yy)f_m(x)$, 
where $f_m(x)$ is determined by the following equation
	\begin{eqnarray}
	-D_m^{xx}f_m''+D_m^{yy}(2\pi Hx/\phi_0)^2f_m+2\omega f_m  \nonumber \\
	+k_z^2(D_m^{zz}-D_m^{xz 2}/D_{m}^{xx})f_m=2\Delta_m,
	\label{uzan}
	\end{eqnarray}
the prime denotes differentiation over x, and $k_y$ was absorbed by the shift of x \cite{ll}. 
The upper critical field $H_{c2}$ is the maximum eigenvalue of Eq. (\ref{uzan}). 
Because of the stability condition $D_m^{zz}D_{m}^{xx}>D_m^{xz 2}$, the term quadratic in $k_z$ 
always decreases the eigenvalues of Eq. (\ref{uzan}), so $H_{c2}$ corresponds to 
$k_z=0$. In this case the lowest eigenfunction $\varphi_0$ of Eq. (\ref{uzan}) is 
	\begin{equation}
	\varphi_0^{(m)}\propto e^{-q_m^2x^2/2}, \qquad q_m=
	\left[\frac{2\pi H}{\phi_0}\right]^{1/2}\!\!\left[\frac{D_m^{yy}}{D_m^{xx}}\right]^{1/4}
	\label{osc}
	\end{equation}
An interesting situation occurs if the ratio $D_m^{yy}/D_m^{xx}$ is different for different bands, as 
characteristic of the $\sigma$ and $\pi$ bands of $MgB_2$. In this case the solution $f_m(x)\propto \Delta_m(x)\propto \varphi_0(x)$
satisfies Eq. (\ref{osc}), but does not satisfy the self-consistency condition (\ref{d}), because $q_1\neq q_2$. This feature 
of two-gap superconductors can essentially complicate the calculation of $H_{c2}$ for inclined fields, as compared to the 
case ${\bf H}\| c$.

$H_{c2}(\theta)$ for an arbitrary field orientation can be obtained by expanding the solution of the 
inhomogeneous Eq. (\ref{osc}) in the series of orthogonal normalized eigenfunctions $\varphi_n(q_mx)$ for   
the different Landau levels n:
	\begin{eqnarray}
	f_m=\sum_{n=0}^{\infty}\frac{\varphi_n(q_mx)}{\omega+(2n+1)\Omega_m}\int_{-\infty}^{\infty}\!\!\Delta_m(u)\varphi_n(q_mu)du, 
	\label{expan} \\
	\varphi_n(q_mx)=\left(\frac{q_m}{\sqrt{\pi}2^nn!}\right)^{1/2}e^{-q_m^2x^2/2}H_n(q_mx),
	\label{eigen} \\
	\Omega_m=(D_m^{xx}D_m^{yy})^{1/2}\pi H/\phi_0, \qquad\qquad
	\label{ome}
	\end{eqnarray}
where $H_n(qx)$ is the Hermitian polynomials \cite{ll}. 
Inserting Eq. (\ref{eigen}) into Eq. (\ref{d}) and summing up over $\omega$, 
gives $H_{c2}$ as the maximum eigenvalue of a matrix $M_{nn'}$, 
for which the determinant $|| M || = 0$. The matrix $M_{nn'}$ is given by 
(see Appendix C):  
	\begin{eqnarray} 
	M_{nn'}=\qquad\qquad\qquad\qquad\qquad
	\nonumber \\
	(1-\lambda_{11}F_n^{(1)})\delta_{nn'}
	-\lambda_{12}\lambda_{21}F_{n'}^{(1)}\sum_{s=0}^{\infty}\frac{V_{ns}V_{n's}F_s^{(2)}}{1-\lambda_{22}F_s^{(2)}},\quad
	\label{matrix} \\
	F_n^{(m)}=\ln\frac{2\gamma\omega_D}{\pi T}-\psi\left(\frac{1}{2}+\frac{\Omega_m(2n+1)}{2\pi T}\right)+\psi\left(\frac{1}{2}\right),\quad 
	\label{fnm} \\
	V_{ns}=\int\varphi_n(q_1x)\varphi_s(q_2x)dx\qquad\qquad\qquad
	\label{vns}
	\end{eqnarray}
where $\delta_{nn'}$ is the Kronecker symbol. 
For ${\bf H}||c$ or for equal anisotropy ratios $D_1^{(a)}/D_1^{(c)}=D_2^{(a)}/D_2^{(c)}$, we have 
$q_1=q_2$, so $V_{nn'}=\delta_{nn'}$, and all off-diagonal terms of the 
matrix M vanish. In this case the equation for the maximum eigenvalue 
$H_{c2}$ is simply $M_{00}=0$, which reduces to Eq. (\ref{hc2}) of the previous paragraph. For arbitrary 
field orientation and general relations between the coupling constants and the anisotropy 
ratios $D_m^{(a)}/D_m^{(c)}$ the equation $||M||=0$ for $H_{c2}$ can 
be solved numerically.  In this case the matrix $M_{nn'}$ should be first diagonalized 
to $\tilde{M}_n\delta_{nn'}$, and then $H_{c2}$ can be found as a root of $\tilde{M}_0=0$.

The equation for $H_{c2}$ greatly simplifies for the moderate anisotropy 
characteristic of $MgB_2$. To quantify the degree of anisotropy, we 
introduce the asymmetry parameter $\zeta=[(q_2^2-q_1^2)/(q_2^2+q_1^2)]^2$, that is, 
	\begin{equation}
	\zeta=\left[\frac{\sqrt{D_2^{xx}/D_2^{yy}} -\sqrt{D_1^{xx}/D_1^{yy}}}
	{\sqrt{D_2^{xx}/D_2^{yy}}+\sqrt{D_1^{xx}/D_1^{yy}}}\right]^2	
	\label{zeta}
	\end{equation}
As shown in the next section, the observed anisotropy of $H_{c2}$ of $MgB_2$ 
near $T_c$ indicates that $D_2^{(a)}\approx D_2^{(c)}$ and $D_1^{(a)}\approx 4D_1^{(c)}$, $\zeta\approx 1/9$.  
In this case the matrix elements $V_{0,2s}\propto \zeta^s$ rapidly decreases  
with $s$ (see Appendix C). By contrast, the factor $F_s^{(2)}/(1-\lambda_{22}F_s^{(2)})$,
which varies much slower with s, can be replaced by its value at $s=0$ and taken out of the sum.  
Indeed, near $T_c$ where $\Omega_m\to 0$, the function  
$F_s\to\ln (2\gamma\omega_D/\pi T)$ is independent of $s$ over a very wide range of s, so 
the above procedure becomes exact. Even for $T=0$, the function 
$F\propto \ln [(2s+1)\Omega_m/T_{c0}]$ varies much weaker as compared to the exponential 
decay of $V_{0s}^2$, so the replacement of $F_s^{(2)}/(1-\lambda_{22}F_s^{(2)})$ by a constant is 
still a very good approximation. The remaining summation over s in Eq. (\ref{matrix}) can then be 
performed exactly, using the sum rule due to the orthogonality of the eigenfunctions $\varphi_n$ (see Appendix C): 
	\begin{equation}
	\sum_{s=0}^{\infty}V_{ns}V_{n's}=\delta_{nn'}
	\label{sumrule}
	\end{equation}
Because of the $\delta_{nn'}$ in Eq. (\ref{sumrule}), all off-diagonal elements of the matrix M vanish, thus the 
equation $\tilde{M}_{00}=0$ for $H_{c2}$ takes the form
	\begin{equation}
	(1-\lambda_{11}F_0^{(1)})(1-\lambda_{22}F_0^{(2)})=\lambda_{12}\lambda_{21}F_0^{(1)}F_0^{(2)}
	\label{foo}
	\end{equation}
In this equation all $q_m(\theta)$ cancel out, and the anisotropy manifests itself 
only via $\Omega_m(\theta)$ in the functions $F_0^{(m)}$ defined by Eqs. (\ref{ome}) and (\ref{fnm}).
After subtraction Eq. (\ref{quadeq}) for $l_c=\ln(2\gamma\omega_D/\pi T_{c0})$, from Eq. (\ref{foo}), the latter reduces to 
Eq. (\ref{hc2}) if the following rescaling $D_m=(D_m^{xx}D_m^{yy})^{1/2}$ is made. 
Therefore, all formulas of the previous section can be generalized to the anisotropic case 
of inclined field by replacing $D_1$ and $D_2$ with the angular-dependent diffusivities 
$D_1(\theta)$ and $D_2(\theta)$ for {\it both bands}:
	\begin{equation}
	D_m(\theta)= [D_m^{(a) 2}\cos^2\theta+D_m^{(a)}D_m^{(c)}\sin^2\theta]^{1/2}
	\label{anis}
	\end{equation}
For example, Eqs. (\ref{hc2}) and (\ref{anis}) give the following 
angular dependence of $H_{c2}(\theta)$ near $T_c$: 	
	\begin{equation}
	H_{c2}(\theta)=\frac{8\phi_0(T_c-T)}{\pi^2[a_1D_1(\theta)+a_2D_2(\theta)]},
	\label{hc2anis} 
	\end{equation} 
For $D_1=D_2$,  Eq. (\ref{hc2anis}) reduces to the well-known result of the anisotropic GL theory, 
$H_{c2}=4\phi_0(T_c-T)/\pi^2D(\theta)$ for a one-gap superconductor in the dirty limit \cite{gork}.

\subsection{Temperature dependence of the anisotropy of $H_{c2}$}

Eqs. (\ref{hc2}) and (\ref{anis}) determine both the temperature and angular dependences of $H_{c2}$. Here the 
anisotropy of $H_{c2}(\theta)$ essentially depends on both T and the in-plane diffusivity ratio $D_2^{(a)}/D_1^{(a)}$. 
For instance, let us consider a simpler case of isotropic diffusivity $D_2^{(a)}=D_2^{(c)}$ for the band 2 and 
anisotropic diffusivity with $D_1^{(a)}\gg D_2^{(c)}$ for the band 1, which qualitatively models the 3D $\pi$ band and the 2D $\sigma$ 
band of $MgB_2$. Then for $D_1^{(a)}\gg D_2^{(ab)}$, the angular dependence of $H_{c2}$ is most pronounced at 
$T_c$ were it is determined by the large anisotropic $D_1(\theta)$ (see Eq. (\ref{hc2anis}), while at lower T the angular dependence 
of $H_{c2}$ decreases, as it is mostly affected by the nearly isotropic small $D_2$ (see Eqs. (\ref{hc2o})-(\ref{o2})). 
However for $D_1^{(a)}\ll D_2^{(a)}$, the situation reverses: the upper critical field is nearly isotropic at $T_c$, 
becoming more anisotropic at lower $T$. These two different types of angular dependencies of $H_{c2}(\theta)$ are 
shown in Fig. 2. 

\begin{figure}          
\epsfxsize= 0.75\hsize
\centerline{
\vbox{
\epsffile{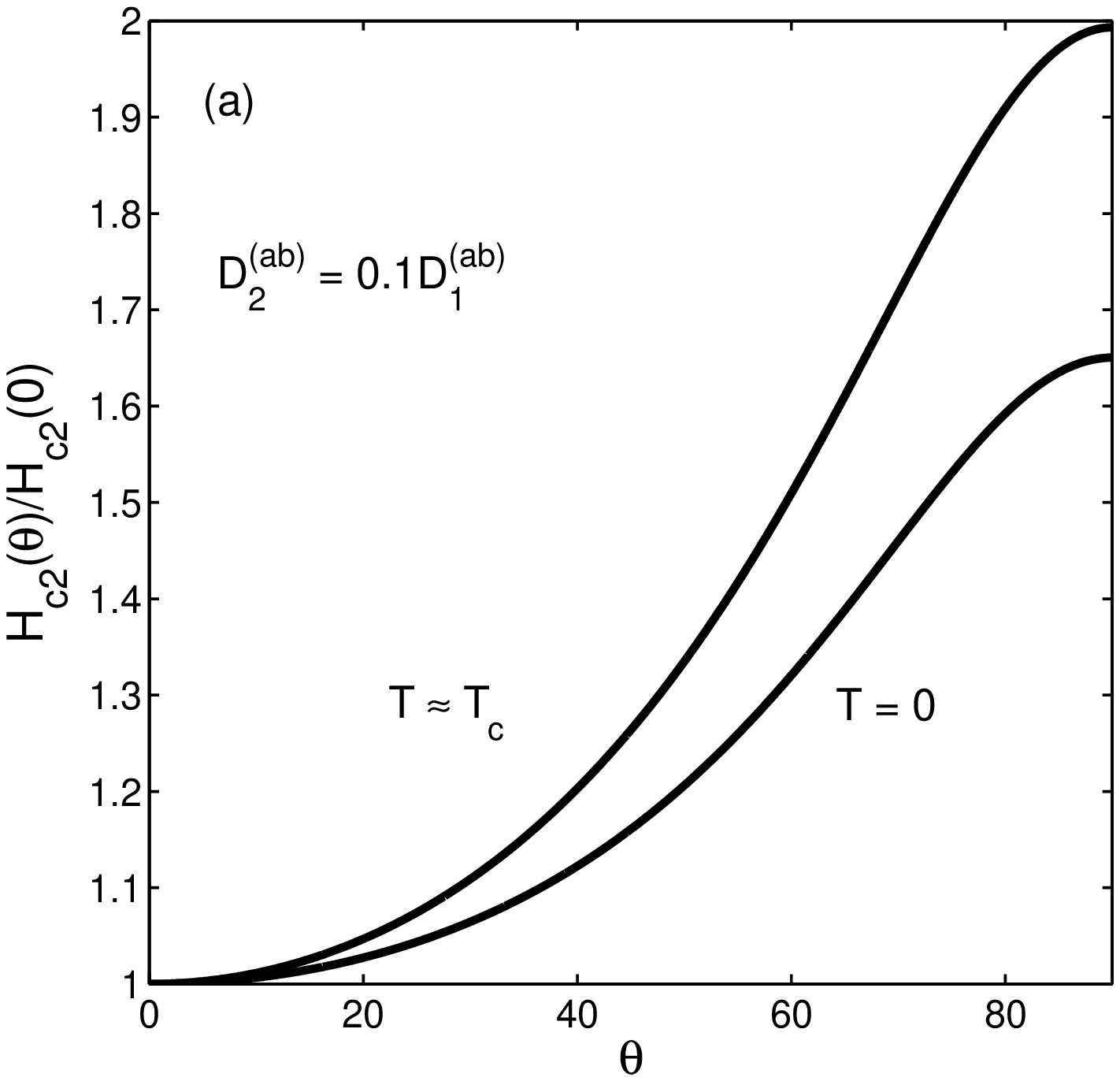}
}}
\epsfxsize= 0.75\hsize
\centerline{
\vbox{
\epsffile{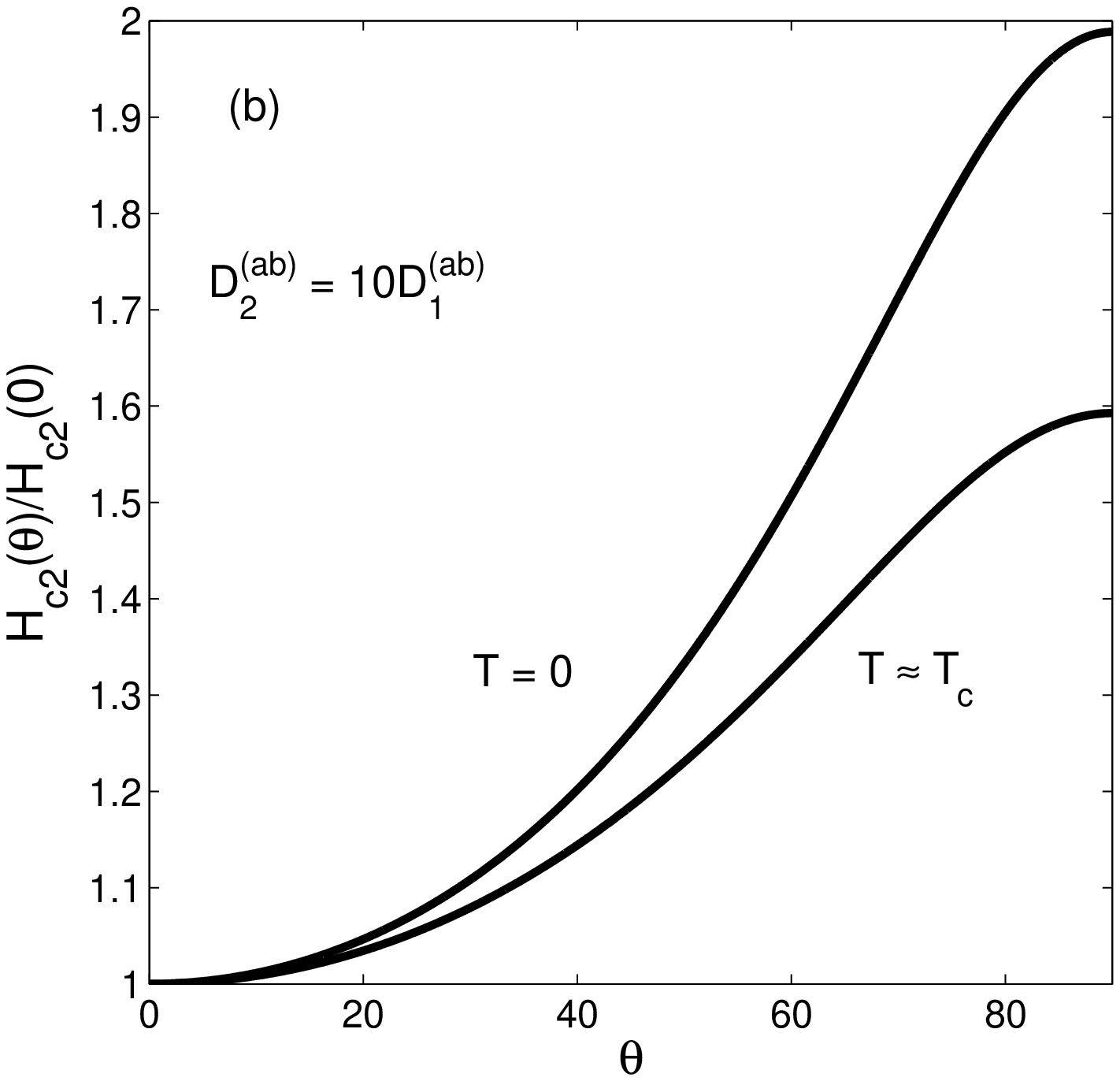}
}}
\vskip \baselineskip
\caption{Angular dependences of $H_{c2}(\theta)$ for $T\approx T_c$ and $T=0$ calculated from Eq. (\ref{hc2anis}) and 
Eq. (\ref{hc2o}), respectively with $D_2^{(c)}=D_2^{(a)}$, $D_1(\theta)=(D_1^{(a)2}\cos^2\theta+D_1^{(c)}D_1^{(a)}\sin^2\theta)^{1/2}$ 
for $D_1^{(a)}=4D_1^{(c)}$. The cases (a) and (b) correspond to different 
ratios $D_2^{(a)}/D_1^{(a)}$, and $\lambda_{mm'}$ are the same as for Fig. 1.
}
\label{fig.2}
\end{figure}

The above arguments indicate that the temperature dependence of the anisotropy parameter 
$H_{c2}^{||}/H_{c2}^{\perp}$ can be rather sensitive to the ratio $D_1/D_2$, which in turn is determined by 
the type of intraband scatterers.  Indeed, if the intraband scattering is dominated by impurities (like C or N) which 
mostly reduce $D_1$ in the main $\sigma$-band, ($D_1\ll D_1$), then, according to Eqs. (\ref{hc2c}) and (\ref{hc2o}), the 
ratio $H_{c2}^{||}/H_{c2}^{\perp}$ is minimum at $T_c$ and increases as T decreases. This behavior has indeed been 
observed on many, mostly clean, $MgB_2$ samples \cite{a1,a2,a3,a4,a5,a6,a7,a8,a9,a10,a11}. 
By contrast, if impurities (like Al, Mg vacancies) cause the strongest intraband scattering in the "weak" $\pi$-band, $(D_2\ll D_1)$, 
then the anisotropy parameter $H_{c2}^{||}/H_{c2}^{\perp}$  would decrease as T  decreases,  The temperature dependence of 
$H_{c2}^{||}/H_{c2}^{\perp}$ for these limiting cases is shown in Fig. 3.  

\begin{figure}          
\epsfxsize= 0.75\hsize
\centerline{
\vbox{
\epsffile{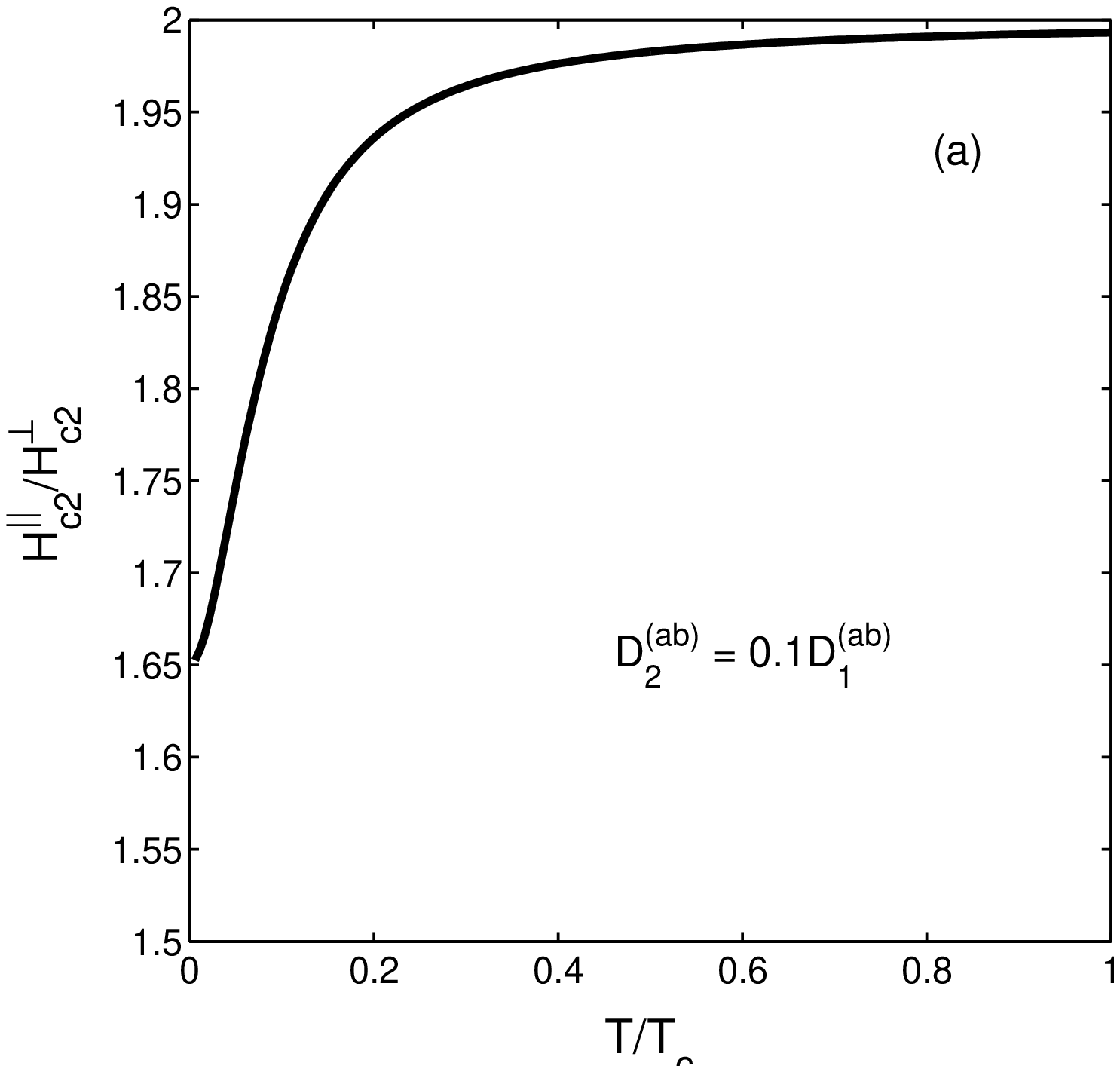}
}}
\epsfxsize= 0.75\hsize
\centerline{
\vbox{
\epsffile{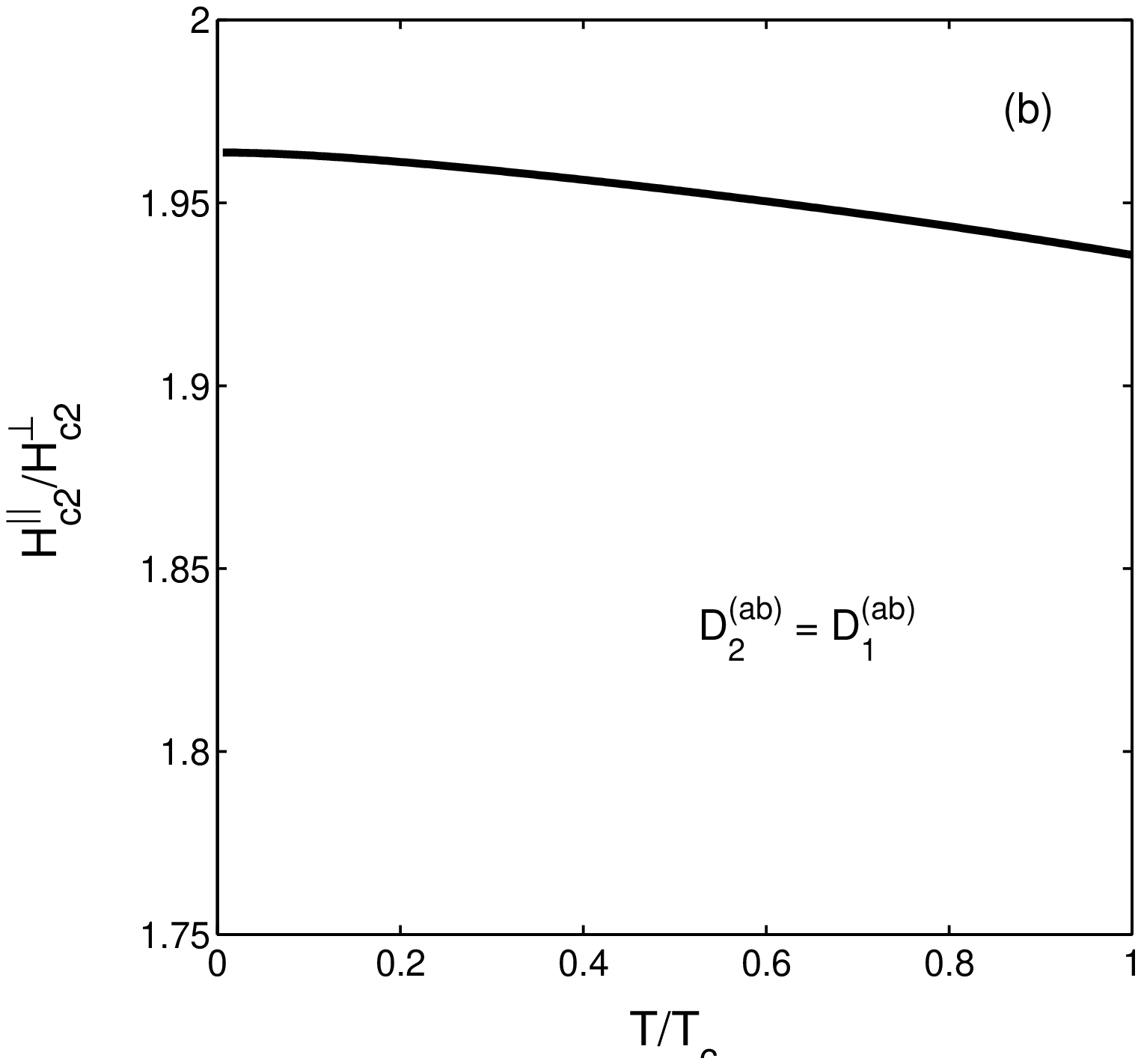}
}}
\epsfxsize= 0.8\hsize
\centerline{
\vbox{
\epsffile{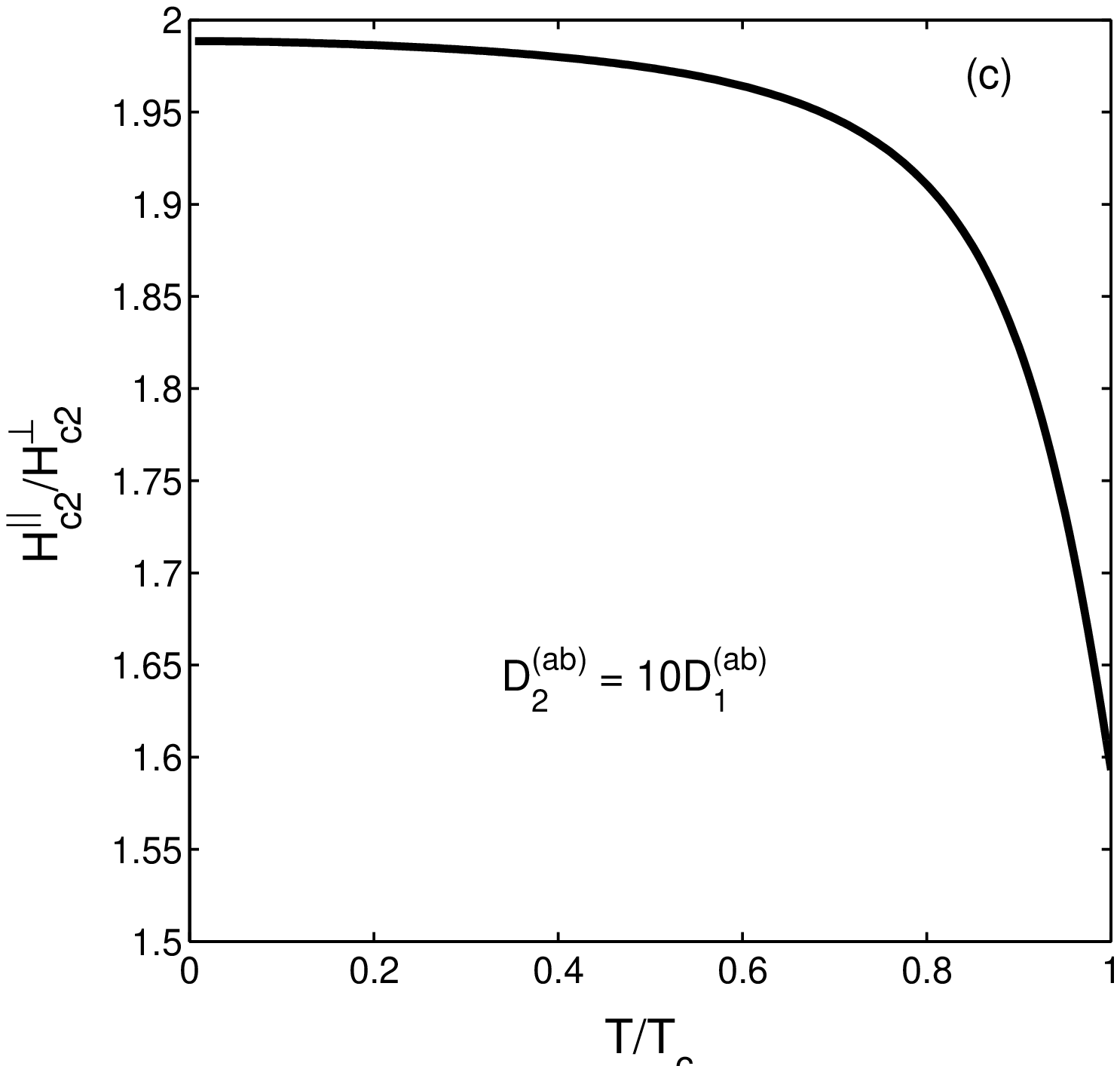}
}}
\vskip \baselineskip
\caption{Temperature dependences of the anisotropy ratio of $H_{c2}^\|/H_{c2}^\perp$ calculated from Eq. (\ref{hc2}) 
with $D_2=D_2^{(a)}$ and $D_1=D_1^{(a)}$ for $H_{c2}^\perp$, and $D_2=D_2^{(a)}=D_2^{(c)}$ and 
$D_1=\sqrt{D_1^{(c)}D_1^{(a)}}$ with $D_1^{(a)}=4D_1^{(c)}$ for $H_{c2}^\|$. The cases (a), (b) and (c) correspond to different 
ratios $D_2^{(a)}/D_1^{(a)}$, and $\lambda_{mm'}$ are the same as for Fig. 1.
}
\label{fig.3}
\end{figure}

\section{Discussion}

The results of this work show that the magnetic behavior of two-gap superconductors can essentially depend on the  
intraband diffusivity ratio $\eta=D_2^{(a)}/D_1^{(a)}$. This fact can have important consequences for $MgB_2$ for which 
$H_{c2}$ can be increased to a much greater extent than in one-band superconductors not just by the usual increase of $\rho_n$, 
but also by tuning the ratio of intraband scattering rates via selective atomic substitutions on both Mg and B sites. 
Actually, this possibility is naturally provided by the electron structure of $MgB_2$ for which substitutions on 
Mg site mostly disturb the out-of-plane $p_z$ orbitals of B thus increasing the $\pi$ scattering. Likewise, atomic 
substitutions on the B sites mostly increase scattering in the 2D $\sigma$ band.  As shown above, the upward curvature 
of $H_{c2}(T)$ for  $D_2^{(a)}\ll D_1^{(a)}$ can make the zero-temperature $H_{c2}(0)$ significantly higher than what the one-gap 
estimate (\ref{whh}) suggests. For example, if $D_\pi = 0.05D_\sigma$ as in Fig. 1a, the field $H_{c2}(0)\approx 2.3T_cH_{c2}'$ 
could approach the paramagnetic limit  of $MgB_2$, $\simeq 70$T for the values $H_{c2}'\approx 1T/K$ and $T_c\approx 30$K 
observed on dirty c-axis oriented $MgB_2$ films\cite{sat}. In fact, recent high-field measurements on this film gave 
$H_{c2}^{\perp}\approx 33.5T$ and $H_{c2}^\|\approx 48T$ which corresponds $D_\pi\approx 0.12D_\sigma$ \cite{gur}. 
Thus, dirty $MgB_2$ can exhibit upper critical fields significantly higher than $H_{c2}(0)\approx 30$T of $Nb_2Sn$ \cite{a15},  
which can be very important for applications \cite{appl}.

The intraband diffusivities $D_1$ and $D_2$ can be either calculated from first principles using Eqs. (\ref{a21}), or extracted 
from measurements of $H_{c2}$ and $\rho_n$. For negligible interband scattering, $D_1$ and $D_2$ can be obtained using 
Eq. (\ref{hc2c}) for $H_{c2}'$ and the relation between intraband diffusivities and the residual conductivity 
$\sigma = 1/\rho_n$:
	\begin{equation}
	\sigma_{\alpha\beta}=e^2\sum_mN_mD_m^{\alpha\beta}
	\label{difsig}
	\end{equation}
For ${\bf H}\| c$, solving two linear equations (\ref{hc2c}) and (\ref{difsig}) expresses $D_1$ and $D_2$ in terms of 
observed $H_{c2}'$ and $\rho_n$:
	\begin{eqnarray}
	D_1=\frac{1}{a_1N_2-a_2N_1}\left(\frac{8k_B\phi_0N_2}{\pi^2\hbar H_{c2}'}-\frac{a_2}{e^2\rho_n}\right),
	\label{d1} \\
	D_2=\frac{1}{a_1N_2-a_2N_1}\left(\frac{a_1}{e^2\rho_n}-\frac{8k_B\phi_0N_1}{\pi^2\hbar H_{c2}'}\right),
	\label{d2}
	\end{eqnarray}
where the normal units are restored.

Using the results of {\it ab-initio} calculations\cite{imp3,goluba} for $MgB_2$, 
$\lambda_{\sigma\sigma}\approx 0.81$, $\lambda_{\pi\pi}\approx 0.285$, $\lambda_{\sigma\pi}\approx 0.119$, 
$\lambda_{\pi\sigma}\approx 0.09$, $N_\sigma\approx 0.3$ states/cell eV, and $N_\pi\approx 0.41$ states/cell eV, 
we obtain $\lambda_-=\lambda_{\sigma\sigma}-\lambda_{\pi\pi}=0.525$, 
$\lambda_0=(\lambda_-^2+4\lambda_{\pi\sigma}\lambda_{\sigma\pi})^{1/2}=0.564$, thus 
$a_1=1+\lambda_-/\lambda_0=1.93$, $a_2=1-\lambda_-/\lambda_0=0.07$, and $a_1N_2\gg a_2N_1$.  
Since both $D_1$ and $D_2$ must be positive, the slope $B_{c2}'$ for a given $\rho_n$ in confined 
within the following limits
	\begin{equation}
	\frac{8ek_B}{\pi a_1}N_1\rho_n< B_{c2}'<\frac{8ek_B}{\pi a_2}N_2\rho_n.
	\label{inequal}
	\end{equation}
For negligible interband coupling $a_2\to 0$, Eq. (\ref{d1}) reduces to the second Eq. (\ref{whh}), 
of the one-gap theory, whereas the stability condition (\ref{inequal}) requires that $B_{c2}'$ cannot be smaller 
than $4ek_BN_1\rho_n/\pi$. This condition gives $B_{c2}'> 0.6 T/K$ for $\rho_n=1 m\Omega$cm, and 
$N_1=0.3$ states/cell eV \cite{imp3}, thus $B_{c2}(0)$ for $T_c=30K$ \cite{sat} cannot be smaller than 
$0.7B_{c2}'T_c = 12.6$T even for  the symmetric case $D_1=D_2$ which gives the minimum 
$H_{c2}(0)$ for a given $H_{c2}'$. For asymmetric case, $D_\pi \ll D_\sigma$ characteristic of 
dirty c-axis oriented $MgB_2$ films \cite{sat}, $H_{c2}(0)$ can considerably exceed this minimum \cite{gur}. 

Another property for which the diffusivity ratio $D_2^{(a)}/D_1^{(a)}$ is essential is   
the temperature dependence of the anisotropy of $H_{c2}$. As shown in the previous section, the  
anisotropy parameter $H_{c2}^{||}/H_{c2}^{\perp}$ increases as T decreases for $D_\sigma\ll D_\pi$, but 
decreases as T decreases for $D_\sigma\gg D_\pi$ (see Fig. 3).  Both cases are in stark contrast with the one-gap theory 
in which the anisotropy ratio for both the upper critical field $H_{c2}^{||}/H_{c2}^{\perp}$ and 
the lower critical field $H_{c1}^{||}/H_{c1}^{\perp}$ is determined by the ratio of the effective masses  
$m_c/m_{ab}$ and is temperature independent. By contrast, the two-gap $MgB_2$ exhibits a strong 
temperature dependence of  $H_{c2}^{||}/H_{c2}^{\perp}$ which is rather different from the 
temperature dependence of $H_{c1}^{||}/H_{c1}^{\perp}$, as was shown both experimentally 
and theoretically\cite{a2,lanis1,lanis2}. For the dirty c-axis oriented films, the observed 
$H_{c2}^{||}/H_{c2}^{\perp}$ is about 2\, \cite{sat}, and $D_\pi\simeq 0.1D_\sigma$\,\cite{gur}. In this case 
the anisotropy of $H_{c2}$ near $T_c$ is entirely determined by the anisotropy of $D_\sigma$, thus 
from Eq. (\ref{hc2anis}) it follows that $D_\sigma^{(a)}\approx 4D_\sigma^{(c)}$.

\section{Acknowlegments}

This work was supported by the NSF  MRSEC (DMR 9214707), and by AFOSR
MURI (49620-01-1-0464). I am grateful to D. C. Larbalestier for illuminating discussions.

\appendix

\section{Derivation of the Uzadel equations}

We derive the Uzadel equations for an anisotropic two-gap superconductor from the  
general Eilenberger equations for the quasiclassical Green functions 
$f({\bf v, r},\omega)$, $f^+({\bf v, r},\omega)$ and $g({\bf v, r},\omega)$ \cite{eilen}
	\begin{eqnarray}
	(2\omega+{\bf v_k\Pi})f({\bf k})&-&2\Delta g(\bf k)= \nonumber \\
	n_i\int\frac{dA_{\bf q}}{v_{\bf q}}|u_{\bf kq}|^2[g({\bf k})f({\bf q})&-&f({\bf k})g({\bf q})],  
	\label{a1}\\
	(2\omega-{\bf v_k\Pi}^*)f^+({\bf k})&-&2\Delta^* g({\bf k})= \nonumber \\
	n_i\int\frac{dA_{\bf q}}{v_{\bf q}}|u_{\bf kq}|^2[g({\bf k})f^+({\bf q})\!&-&\!f^+({\bf k})g({\bf q})].  
	\label{a2}
	\end{eqnarray}
Here $dA_{\bf k}/v_{\bf k}$ means integration over the FS with the local density of states 
$1/v_{\bf k}$, the integral terms account for nonmagnetic impurity scattering, 
$u_{\bf k,q}$ is the scattering amplitude, $n_i$ is the density of impurities, $v_{\bf k}$ is the normal group velocity 
on an anisotropic FS, the wave vectors ${\bf k}$ and $\bf{q}$ lie on the FS, and the asterisk means complex conjugate.  
Eqs. (\ref{a1}) and (\ref{a2}) are supplemented by the normalization condition
	\begin{equation}
	ff^++|g|^2=1,
	\label{a3}
	\end{equation}
the equation for the superconducting gap
	\begin{equation}
	\Delta({\bf k, r})=2\pi T\sum_{\omega>0}^{\omega_D}\int\frac{dA_{\bf q}}{v_{\bf q}}V({\bf k, q})f({\bf q, r},\omega),
  	\label{a4}
	\end{equation}
and the expression for the current density
	\begin{equation}
	{\bf J}=-4\pi Te Im\sum_{\omega>0}^{\omega_D}\int\frac{dA_{\bf q}}{v_{\bf q}}{\bf v_q}g({\bf q, r},\omega),
	\label{a5}
	\end{equation}
where $V({\bf k,q})$ is the pairing potential.

Now we define the Green functions $f_1$ and $f_2$ on two separate sheets of the FS, and  
write the collision integrals in the right hand side of Eqs. ({\ref{a1}) and (\ref{a2}) in the form
	\begin{eqnarray} 
            St_1=n_i\int\frac{dA_{\bf q_1}}{v_{\bf q_1}}|u_{\bf k_1q_1}|^2[g_1({\bf k_1})f_1({\bf q_1})-f_1({\bf k_1})g_1({\bf q_1})] 
	\nonumber \\  
	+ n_i\int\frac{dA_{\bf q_2}}{v_{\bf q_2}}|u_{\bf k_1q_2}|^2[g_1({\bf k_1})f_2({\bf q_2})-f_1({\bf k_1})g_2({\bf q_2})]\qquad  
	\label{a6}
	\end{eqnarray}
	\begin{eqnarray} 
            St_2=n_i\int\frac{dA_{\bf q_2}}{v_{\bf q_2}}|u_{\bf k_2q_2}|^2[g_2({\bf k_2})f_2({\bf q_2})-f_2({\bf k_2})g_2({\bf q_2})] 
	\nonumber \\  
	+n_i\int\frac{dA_{\bf q_1}}{v_{\bf q_1}}|u_{\bf k_2q_1}|^2[g_2({\bf k_2})f_1({\bf q_1})-f_2({\bf k_2})g_1({\bf q_1})]\qquad  
	\label{a7}
	\end{eqnarray}
The first integral term in Eqs. (\ref{a6}) and (\ref{a7}) describe the {\it intraband} scattering for which the wave vectors 
${\bf k}$ and ${\bf q}$ lie on the same sheet (1 or 2) of the FS, while the second integral term 
describes the {\it interband} scattering for which ${\bf k}$ and ${\bf q}$ lie on the different FS sheets. 

To derive the equations for $f_m({\bf r},\omega)$, $f^+_m({\bf k},\omega)$, $g_m({\bf k, r},\omega)$, 
and $\Delta_m({\bf r})$ in the dirty limit, we follow the procedure developed for one-band superconductors, 
expanding $f({\bf k, r},\omega)$ and $g({\bf k, r},\omega)$, in spherical harmonics, and keeping only the 
first dipole term \cite{uzadel}
	\begin{eqnarray}
	f_m({\bf k, r},\omega)=f_m({\bf r},\omega)+({\bf v}_m\delta{\bf f}_m({\bf r},\omega)),
	\label{a8} \\
	g_m({\bf k, r},\omega)=g_m({\bf r},\omega)+({\bf v}_m\delta{\bf g}_m({\bf r},\omega)),
	\label{a9}
	\end{eqnarray}
where the band index $m$ runs from 1 to 2, and the last terms in the right hand side describe 
first order anisotropic corrections to $f({\bf k, r},\omega)$ and $g({\bf k, r},\omega)$. These corrections 
are small if $6\pi T_c\gg v_m^2/D_m$, where $D_m$ are intraband diffusivities (see below). 
The vector corrections $\delta{\bf g}_m$ and $\delta{\bf f}_m$ are related through the normalization condition (\ref{a3}):  
	\begin{equation}
	2g_m\delta{\bf g}_m = - f_m^+\delta{\bf f}_m-f_m\delta{\bf f}_m^+
	\label{a10}
	\end{equation}
Now we insert Eqs. (\ref{a8}) and (\ref{a9}) into Eqs. (\ref{a1})-(\ref{a2}) and (\ref{a6})-(\ref{a7}) and 
integrate over the respective FS sheets: $\int dA_{\bf k}/v_m({\bf k})$, using  
$\int {\bf v}_m({\bf k})dA_{\bf k}=0$. Since all $f_m$ and $\delta{\bf f}_m$ are independent of ${\bf v}$, this yields
	\begin{eqnarray}
	2\omega f_1-2\Delta_1g_1+r_1^{\alpha\beta}\Pi_\alpha\delta f_1^\beta=(g_1f_2-g_2f_1)\gamma_{12},
	\label{a11} \\
	2\omega f_2-2\Delta_2g_2+r_2^{\alpha\beta}\Pi_\alpha\delta f_2^\beta=(g_2f_1-g_1f_2)\gamma_{21},
	\label{a12} \\
	r_m^{\alpha\beta}=\int v_m^\alpha({\bf k}) v_m^\beta({\bf k}) \frac{dA_{\bf k}}{v_m({\bf k})N_m}\qquad\qquad
	\end{eqnarray}
where $N_m=\int dA_{\bf k}/v_m({\bf k})$ are partial densities of states, and $\gamma_{mm'}$ are 
the interband scattering rates \cite{allen}
	\begin{equation}
	\gamma_{mm'}=\frac{n_i}{N_m}\int \frac{dA_{\bf q}}{v_m({\bf q})}\frac{dA_{\bf k}}{v_{m'}({\bf k})}|u_{{\bf k,q}}|^2
	\label{a13}
	\end{equation}
Then we substitute Eqs. (\ref{a8}) and (\ref{a9}) into Eqs. (\ref{a1})-(\ref{a2}) and (\ref{a6})-(\ref{a7}), 
multiply them by ${\bf v}_m({\bf k})$ and integrate over the respective FS sheets, using the fact that 
$\int {\bf v}_1({\bf k}){\bf v}_2({\bf k})dA_{\bf k}=0$. In the dirty limit, this yields the following relations
	\begin{eqnarray}
	\Gamma_m^{\alpha\beta}(f_m\delta g_m^\beta-g_m\delta f_m^\beta)=r_m^{\alpha\beta}\Pi_\beta f_m, 
	\label{a14} \\
	\Gamma_m^{\alpha\beta}(f_m^+\delta g_m^\beta-g_m\delta f_m^{+\beta})=-r_m^{\alpha\beta}\Pi_\beta^* f_m^+,
	\label{a15} 
	\end{eqnarray} 
Here the tensor $\Gamma^{\alpha\beta}_m$ is defined by 
	\begin{equation}
	\Gamma^{\alpha\beta}_m=\frac{n_i}{N_m}\int
	\frac{|u_{{\bf q, k}}|^2v^\alpha_m({\bf k})[v^\beta_m({\bf k})-v^\beta_m({\bf q})]}{v_m({\bf k})v_{m}({\bf q})}dA_{\bf k}dA_{\bf q}
	\label{a16}
	\end{equation}
Eqs. (\ref{a14}) and (\ref{a15}) can be rewritten as 
 	\begin{eqnarray}
	f_m\delta g_m^\alpha-g_m\delta f_m^\alpha=F_m^{\alpha\beta}\Pi_\beta f_m, 
	\label{a17} \\
	f_m^+\delta g_m^\alpha-g_m\delta f_m^{+\alpha}=-F_m^{\alpha\beta}\Pi_\beta^* f_m^+,
	\label{a18} 
	\end{eqnarray} 
where $F_m^{\alpha\beta}=(\Gamma_m^{\alpha\gamma})^{-1}r_m^{\gamma\beta}$. 
Multiplying Eq. (\ref{a17}) by $f_m^+$ and Eq. (\ref{a18}) by $f_m$, adding, and then using 
Eq. (\ref{a10}), we obtain
	\begin{equation}
	2\delta g_m^\alpha=F_m^{\alpha\beta}(f_m^+\Pi_\beta f_m-f_m\Pi_\beta^*f_m^+)
	\label{a19}
	\end{equation}
Using Eq. (\ref{a14}), (\ref{a15}), (\ref{a19}), and the normalization condition $g_m^2+|f_m|^2=1$, we obtain
	\begin{equation}
	\delta f_m^\alpha=-F_m^{\alpha\beta}(g_m\Pi_\beta f_m-f_m\nabla_\beta g_m).
	\label{a20}
	\end{equation} 
Inserting Eq. (\ref{a20}) and its conjugate into Eqs. (\ref{a11}) and (\ref{a12}), we arrive at Eqs. (\ref{uz1}) 
and (\ref{uz2}) in which the intraband diffusivity tensors are given by 
	\begin{equation}
	D_m^{\alpha\beta}=r_m^{\alpha\gamma}(\Gamma_m^{\gamma\mu})^{-1}r_m^{\mu\beta}.
	\label{a21}
	\end{equation}
Averaging the gap equation (\ref{a4}) over the FS, yields Eq. (\ref{d}) in which the coupling constants $\lambda_{mm'}$ are given by
	\begin{equation}
	\lambda_{mm'}=\frac{1}{N_m}\int\frac{dA_{\bf k}dA_{\bf q}}{v_m({\bf k})v_{m'}({\bf q})}V({\bf k},{\bf q}),
	\label{a22}
	\end{equation} 
whence $N_1\lambda_{12}=N_2\lambda_{21}$.

\section{Equations for $H_{c2}$ and $T_c$ with the account of interband scattering}

For $H||c$, the linearized Eqs. (\ref{uz1}) and (\ref{uz2}) take the form
	\begin{eqnarray}
	\omega f_1-\frac{D_1}{2}\Pi^2f_1=\Delta_1+(f_2-f_1)\gamma_{12},
	\label{b1} \\
	\omega f_2-\frac{D_2}{2}\Pi^2f_2=\Delta_2+(f_1-f_2)\gamma_{21},
	\label{b2}
	\end{eqnarray}
In the gauge $A_y=Hx$, the solution of Eqs. (\ref{b1}) and (\ref{b2}) is
	\begin{equation}
	f_m(x)=\tilde{f}_me^{-q^2x^2/2}, \qquad \Delta_m(x)=\tilde{\Delta}_me^{-q^2x^2/2}.
	\label{b3}
	\end{equation}
Here $q^2=2\pi H/\phi_0$, and the amplitudes $\tilde{f}_m$ can be expressed via $\tilde{\Delta}_1$ and 
$\tilde{\Delta}_2$ from Eqs. (\ref{b1})-(\ref{b3}):
	\begin{eqnarray}
	\tilde{f}_1=\frac{g_+\tilde{\Delta}_1-g_{12}\tilde{\Delta}_2}{\omega+\Omega_+}+
	\frac{g_-\tilde{\Delta}_1+g_{12}\tilde{\Delta}_2}{\omega+\Omega_-},
	\label{b4} \\
	\tilde{f}_2=\frac{g_-\tilde{\Delta}_2-g_{21}\tilde{\Delta}_1}{\omega+\Omega_+}+
	\frac{g_+\tilde{\Delta}_2+g_{21}\tilde{\Delta}_1}{\omega+\Omega_-},
	\label{b5}
	\end{eqnarray}
where 
	\begin{eqnarray}
	2\Omega_+ =\omega_++\gamma_++\Omega_0,\qquad 2\Omega_-=\omega_++\gamma_+-\Omega_0,\quad
	\label{b6} \\
	\Omega_0=(\omega_-^2+\gamma_+^2+2\gamma_-\omega_-)^{1/2}, \qquad\qquad\quad
	\label{b7} \\
	\gamma_\pm=\gamma_{12}\pm\gamma_{21}, \qquad \omega_\pm=(D_1\pm D_2)\pi H/\phi_0,\qquad
	\label{b8} \\
	2g_\pm=1\pm\frac{\omega_-+\gamma_-}{\Omega_0},
	\quad g_{12}=\frac{\gamma_{12}}{\Omega_0}, \quad g_{21}=\frac{\gamma_{21}}{\Omega_0}.\qquad
	\label{b9}
	\end{eqnarray}
Inserting Eqs. (\ref{b4}) and (\ref{b5}) into Eq. (\ref{d}) we first sum up over $\omega$, 
and express the result in terms of the functions 
	\begin{eqnarray}
	F_\pm=2\pi T\sum_\omega\frac{1}{\omega+\Omega_\pm}=\ln\frac{2\gamma\omega_D}{\pi T}-
	\nonumber \\ 
	-\psi\left(\frac{1}{2}+\frac{\Omega_\pm}{2\pi T}\right)+\psi\left(\frac{1}{2}\right).
	\label{b10}
	\end{eqnarray}
Then the self-consistency equations (\ref{d}) reduce to
	\begin{eqnarray}
	G_{11}{\tilde\Delta}_1+G_{12}{\tilde\Delta}_2=0, 
	\label{b11} \\
	G_{21}{\tilde\Delta}_1+G_{21}{\tilde\Delta}_2=0,
	\label{b12}
	\end{eqnarray}
where the matrix $G_{mm'}$ is given by
	\begin{eqnarray}
	G_{11}=1-\lambda_{11}(g_+F_++g_-F_-)+\lambda_{12}g_{21}(F_+-F_-),\quad
	\label{b13} \\
	G_{12}=\lambda_{11}g_{12}(F_+-F_-)-\lambda_{12}(g_-F_++g_+F_-),\qquad
	\label{b14} \\
	G_{22}=1-\lambda_{22}(g_+F_++g_-F_-)+\lambda_{21}g_{12}(F_+-F_-),\quad
	\label{b15} \\
	G_{21}=\lambda_{22}g_{21}(F_+-F_-)-\lambda_{21}(g_+F_++g_-F_-).\qquad
	\label{b16}
	\end{eqnarray}
Using Eqs. (\ref{b13})-(\ref{b16}), the final equation $G_{11}G_{22}-G_{12}G_{21}=0$ for $H_{c2}$ 
can be presented in the form
	\begin{eqnarray}
	1-\lambda_{11}(g_+F_++g_-F_-)- \lambda_{22}(g_+F_++g_-F_-)\quad
	\nonumber \\
	+w(g_+F_++g_-F_-)(g_+F_++g_-F_-)\qquad
	\nonumber \\
	+(\lambda_{12}g_{21}+\lambda_{21}g_{12})(F_+-F_-)=0,\qquad\qquad
	\label{b17}
	\end{eqnarray}
where $H_{c2}$ is the maximum root of Eq. (\ref{b17}).
Setting $H_{c2}=0$ in Eq. (\ref{b17}), gives the equation for $T_c$ with the account of interband scattering 
for which $\omega_\pm=\Omega_-=0$, $\Omega_+=\gamma_{12}+\gamma_{21}$, 
$g_+=2\gamma_{12}/(\gamma_{12}+\gamma_{21})$, $g_-=2\gamma_{21}/(\gamma_{12}+\gamma_{21})$. 
Subtracting Eq. (\ref{quadeq}) for $T_{c0}$ from Eq. (\ref{b17}), the equation for $T_c$ can be reduced to
	\begin{eqnarray}
	U\left(\frac{\gamma_{12}+\gamma_{21}}{2\pi T}\right)=\frac{(\lambda_0-w\ln t)\ln t}{p-w\ln t},\qquad
	\label{b18} \\
	p=\frac{\gamma_{12}(\lambda_1 -2\lambda_{21})+\gamma_{21}(\lambda_2-2\lambda_{12})}
	{2(\gamma_{12}+\gamma_{21})},\quad
	\label{b19}
	\end{eqnarray}
where $t=T_c/T_{c0}$, $\lambda_1=\lambda_0+\lambda_-$, and $\lambda_2=\lambda_0-\lambda_-$.
The effect of interband scattering on $T_c$ suppression was considered before\cite{imp1,imp2,lanis1}

\section{Matrix M}

Inserting Eq. (\ref{expan}) into the gap equation (\ref{d}) yields  
	\begin{equation}
	\Delta_m=2\pi T\sum_\omega\sum_{m'}\lambda_{mm'}\sum_{n=0}^{\infty}\frac{\varphi_n(q_{m'}x)C_n^{(m')}}
	{\omega+(2n+1)\Omega_{m'}},
	\label{c1}
	\end{equation}
where $C_n^{(m)}=\int_{-\infty}^{\infty}\varphi_n(q_mx)\Delta_m(x)dx$. 
Now we multiply Eq. (\ref{c1}) by $\varphi_s(q_mx)$, integrate over x 
using the orthogonality condition $\int\varphi_n(q_mx)\varphi_{n'}(q_mx)dx=\delta_{nn'}$, and then sum up over $\omega$. 
This gives two sets of linear equations 
	\begin{eqnarray}
	(1-\lambda_{11}F_s^{(1)})C_s^{(1)}=\lambda_{12}\sum_nV_{sn}F_n^{(2)}C_n^{(2)}, 
	\label{c2} \\
	(1-\lambda_{22}F_s^{(2)})C_s^{(2)}=\lambda_{21}\sum_nV_{ns}F_n^{(1)}C_n^{(1)}, 
	\label{c3}
	\end{eqnarray}
where $V_{sn}$ and $F_n^{(m)}$ are defined as follows
	\begin{eqnarray}
	V_{sn}=\int_{-\infty}^{\infty}\varphi_s(q_1x)\varphi_n(q_2x)dx
	\label{c4} \\
	F_n^{(m)}=\sum_{\omega > 0}^{\omega_D}\frac{1}{\omega+(2n+1)\Omega_{m}}
	\label{c5}
	\end{eqnarray}
Eq. (\ref{c5}) can be expressed in terms of the $\psi$ function as 
in Eq. (\ref{fnm}). Inserting $C_n^{(2)}$ from Eq. (\ref{c3}) into Eq. (\ref{c2}) yields the closed set 
of linear equations for $C_n^{(1)}$ 
	\begin{equation}
	\sum_{n'=0}^{\infty}M_{nn'}C_{n'}^{(1)}=0,
	\label{c6}
	\end{equation}
where the matrix $M_{nn'}$ is given by Eq. (\ref{matrix}). The solvability condition of Eq. (\ref{c6}) 
gives the equation for $H_{c2}$, which is the maximum eigenvalue of the matrix M.

A general formula for $V_{nn'}$ is rather cumbersome, so to illustrate general behavior of $V_{nn'}$ we consider the 
case of $n'=0$ for which the integral (\ref{c4}) equals \cite{ryzik}
	\begin{equation}
	V_{n0}=\left(\frac{2q_1q_2(2n)!}{2^{2n}(n!)^2(q_1^2+q_2^2)}\right)^{1/2}\left(\frac{q_2^2-q_1^2}{q_2^2+q_1^2}\right)^{n},
	\label{c7} 
	\end{equation}
if $n=2s$, and $V_{n0}=0$ for $n=2s+1$, where s is any positive integer. As follows from Eq. (\ref{c7}), 
$V_{2s,0}\propto \zeta^s$ exponentially decays with s.  The matrix elements $V_{nn'}$ obey the useful sum rule 
$\sum_{s=0}^{\infty}V_{ns}V_{n's}=\delta_{nn'}$, which can be proven by inserting Eq. (\ref{c4}) under the sum:
	\begin{equation}
	\sum_{s=0}^{\infty}\int_{-\infty}^{\infty}dxdx'\varphi_n(q_1x)\varphi_s(q_2x)\varphi_{n'}(q_1x')\varphi_s(q_2x')
	\label{c8}
	\end{equation}
To sum up over s, we use the identity \cite{ll}
	\begin{equation}
	\sum_{s=0}^{\infty}\varphi_s(q_2x)\varphi_s(q_2x')=\delta(x-x')
	\label{c9}
	\end{equation}
After inserting Eq. (\ref{c9}) into Eq. (\ref{c8}), the latter reduces to $\int\varphi_n(q_1x)\varphi_{n'}(q_1x)dx$, 
giving the sum rule (\ref{sumrule}).

Another equation for $H_{c2}$ valid  for weak interband coupling $\lambda_{12}\lambda_{21}\ll \lambda_{11}\lambda_{22}$ and any    
anisotropy can be obtained as follows. Since off-diagonal terms of M in Eq. (\ref{matrix}) are proportional to the small factor 
$\lambda_{12}\lambda_{21}$, the matrix element $\tilde{M}_0$ can be obtained in the second order perturbation theory: 
	\begin{equation}
	\tilde{M}_0=M_{00}-\sum_{s=1}^{\infty}\frac{M_{0,2s}M_{2s,0}}{M_{2s,2s}}
	\label{c10}
	\end{equation}	
Here $M_{0,2s}$ is proportional to the small parameter $\lambda_{12}\lambda_{21}\zeta$, so  
the sum in Eq. (\ref{c10}) is by the factor $\sim \lambda_{12}\lambda_{21}\zeta^2$ smaller  
than the sum in the right hand side of Eq. (\ref{matrix}) for $M_{00}$. Thus, the contribution of the off-diagonal terms to 
$\tilde{M}_0$ can be neglected, and the equation for $H_{c2}$ is again $M_{00}=0$, that is,
	\begin{equation}
	1-\lambda_{11}F_0^{(1)}=
	\lambda_{12}\lambda_{21}F_0^{(1)}\sum_{s=0}^{\infty}\frac{V_{0s}^2F_s^{(2)}}{1-\lambda_{22}F_s^{(2)}}
	\label{c11} 
	\end{equation}
For moderate anisotropy $(\zeta \ll 1)$, Eq. (\ref{c11}) reduces to Eq. (\ref{foo}), as the convergence of the sum in 
Eq. (\ref{c11}) is due to the rapidly decreasing factor $V_{0s}^2\propto \zeta^s$.  However, for very strong anisotropy 
$\zeta\approx 1$ at low T when the dependence of $F_s^{(2)}$ on s 
is essential, Eq. (\ref{c11}) may be simpler for numerical solution than the full Eq. (\ref{c6}).


\begin{references}

\bibitem{tg1}
A. Liu, I.I. Mazin, and J. Kortus, \prl {\bf 87} 087005 (2001). 							

\bibitem{tg2}
H.J. Choi, D. Roundy, H. Sun, M.L. Cohen, and S.G. Loule, Nature {\bf 418}, 758 (2002);			
\prb {\bf 66}, R020513 (2002).

\bibitem{tg3}
G. Karapetrov, M. Iavarone, W.K. Kwock, G.W. Crabtree, and D.G. Hinks, \prl {\bf 86}, 4374 (2001); 
M. Iavarone, et al. {\it ibid.}  {\bf 89}, 1870021 (2002). 							

\bibitem{tg4}
M.R. Esklidsen, M. Kugler, S. Tanaka, J. Jun, S.M. Kazakov, J. Karpinski, and O. Fisher, 
\prl {\bf 89}, 197003 (2002).											

\bibitem{tg5}
H. Schmidt, J.E. Zasadzinski, K.E. Gray, and D.G. Hinks, D.G. \prl {\bf 88}, 1270021-4 (2002). 		

\bibitem{tg6}
X.K. Chen, M.J. Kostantinovic, J.C. Irwin, D.D. Lawrie, and J.P. Franck, \prl {\bf 87}, 1570021 (2001). 	

\bibitem{tg7}
F. Bouquet, R.A. Fisher, N.E. Phillips, D.G. Hinks, and J.D. Jorgensen, J.D. \prl {\bf 87}, 047001 (2001). 	

\bibitem{tg8}
J.J. Tu, {\it et al.}, \prl {\bf 87}, 277001 (2001); 
M.-S. Kim, {\it et al.}, \prb {\bf 66}, 0645011 (2002); 								
A.B. Kuz'menko {\it et al.}, Solid State Comm. {\bf 121}, 479 (2002).						

\bibitem{rev}	
C. Buzea and T. Yamashita, Supercond. Sci. Technol. {\bf 14}, R115 (2001).				

\bibitem{appl}													
P. Grant, Nature {\bf 411}, 532 (2001);
D. Larbalestier, A. Gurevich, D.M. Feldmann, and A. Polyanskii, {\it ibid.} {\bf 414}, 368 (2001). 

\bibitem{a1}
O.F. de Lima, R.A. Ribeiro, M.A. Avila, C.A. Cardoso, and A.A. Coelho, \prl {\bf 87}, 047002 (2001).

\bibitem{a2}
S.L. Bud'ko, V.G. Kogan, and P.C. Canfield, \prb {\bf 64}, 180506 (2001); S.L. Bud'ko and P.C. Canfield, 
{\it ibid.}, {\bf 65}, 212501 (2002).

\bibitem{a3}
S. Lee, H. Mori, T. Masui, Y. Eltsev, A. Yamamoto, and S. Tajima, J. Phys. Soc. Jap. {\bf 70}, 2255 (2001).

\bibitem{a4}
A.K. Pradhan, Z.X. Shi, M. Tokunaga, T. Tamegai, Y. Takano, K. Togano, M. Kito, and H. Ihara, 
\prb {\bf 64}, 212509 (2001). 

\bibitem{a5}
M. Xu, H. Kitazawa, Y. Takano, J. Ye, K. Nishava, H. Abe, A. Matsushita, N. Tsujiiand G. Kido, 
Appl. Phys. Lett. {\bf 79}, 2779 (2001).

\bibitem{a6}
M. Angst, R. Puzniak, A. Wisiewski, J. Jun, S.M. Kazakov, J. Karpinski, J.Ross, and H. Keller, 
\prl {\bf 88}, 167004 (2002). 

\bibitem{a7}
A.V. Sologubenko, J. Jun, S.M. Kazakov, J. Karpinski, and H.R. Ott, \prb {\bf 65}, 180505 (2002).

\bibitem{a8}
K. Takahashi, T. Atsumi, N. Yamamoto, M. Xu, H. Kitazawa, and T. Ishida, \prb {\bf 66}, 012501 (2002).

\bibitem{a9}
M. Zehetmayer, M. Eisterer, J. Jun, S.M. Kazakov, J. Karpinski, A. Wisniewski, and H.W. Weber, 
\prb {66}, 052505 (2002).

\bibitem{a10}
L. Lyard {\it et al.,} cond-mat/0206231.

\bibitem{a11}
U. Welp {\it et al.}, cond-mat/0203337. 

\bibitem{whh}
P.G. De Gennes, Phys. Cond. Materie {\bf 3}, 79 (1964); 
N.R. Werthamer, E. Helfand, and P.C. Hohenberg, Phys. Rev. {\bf 147}, 295 (1966); 
K. Maki, {\it ibid.}, {\bf 148}, 362 (1966).

\bibitem{a15}
W.A. Fietz and W.W. Webb, Phys. Rev. {\bf 161}, 4231 (1967);
T.P. Orlando, E.J. McNiff, S. Foner, and M.R. Beasley, \prb {\bf 19}, 4545 (1979).

\bibitem{perk}
G.K. Perkins, G.K., et al. Nature, 411, 561-563 (2001).			

\bibitem{buket}
M. Eisterer {\it et al.,} Supercond. Sci. Technol. {\bf 15}, L9 (2002);
Y. Wang {\it et al.,} cond-mat/0208169 					

\bibitem{sat}
S. Patnaik {\it et al.}, Supercond. Sci. Technol. {\bf 14}, 315 (2001). 

\bibitem{jc1}
A. G{\" u}mbel, J. Eckert, G. Fuchs, K. Nenkov, K.H. M{\" u}ller, and L. Schultz, 
Appl. Phys. Lett. {\bf 80}, 2725 (2002).

\bibitem{jc2}
K. Komori, K. Kawagishi, Y. Takano, H. Fujii, S. Arisawa, H. Kumakura, M. Fukutomi, and K. Togano,
Appl. Phys. Lett. {\bf 81}, 1047 (2002).

\bibitem{fin}
D.K. Finnemore, J.L. Ostenson, S.L. Bud'ko, G. Lapertot, and P.C. Canfield, \prl {\bf 86}, 2420 (2001).

\bibitem{gur}
A. Gurevich, et al, cond-mat/

\bibitem{suhl}
H. Suhl, B.T. Matthias, and L.R. Walker, \prl {\bf 12}, 552 (1959).

\bibitem{imp1}
S.V. Pokrovsky and V.L. Pokrovsky, \prb{\bf 54}, 13275 (1996); 

\bibitem{imp2}
A.A. Golubov and I.I. Mazin, \prb {\bf 55}, 15146 (1997).

\bibitem{imp3}
I.I. Mazin {\it et al.}, \prl {\bf 89}, 107002 (2002).

\bibitem{lanis1}
V.G. Kogan, \prb {\bf 66}, R020509 (2002).					

\bibitem{lanis2}
A.A. Golubov, A. Brinkman, O.V. Dolgov, J. Kortus, and O. Jepsen, 		
\prb {bf 66}, 054524 (2002).

\bibitem{goluba}
A.A. Golubov, J. Kortus, O.V. Dolgov, O. Jepsen, Y. Kong, O.K. Andersen, B.J. Gibson, K. Ahn, and R.K. Kremer, 
J. Phys: Condens. Matt. {\bf 14}, 1353 (2002).h

\bibitem{shulga}
S.V. Shulga {\it et al.}, \prl {\bf 80}, 1730 (1998); cond-mat/0103154.

\bibitem{eilen}
G. Eilenberger, Z. Phys. {\bf 214}, 195 (1969); A.I. Larkin and 
Yu.N. Ovchinnikov, Zh. Exp. Teor. Fiz. {\bf 55}, 2262 (1968) 
[Sov. Phys. JETP {\bf 28}, 1200 (1969)]. 

\bibitem{book}
J.B. Ketterson and S.N. Song, {\it Superconductivity} (Cambridge University Press, Cambridge, New York, 1999).

\bibitem{uzadel}
K. Uzadel, \prl {\bf 25}, 507 (1970).

\bibitem{allen} 
P.B. Allen, \prb {\bf 13}, 1416 (1976).

\bibitem{koshgol}
A.E. Koshelev and A.A. Golubov, cond-mat/0211388. In this work an equation for $H_{c2}$   
equivalent to Eq. (\ref{hc2}) divided by $[\ln t + U(h)][\ln t + U(\eta h)]$ was obtained. After such division 
however this equation does not reproduce the correct result, $\ln t + U(h)=0$, for $D_1=D_2$. 

\bibitem{tanaka}
Y. Tanaka, \prl {\bf 88}, 017002 (2002); E. Babaev {\it ibid.}, {\bf 89}, 067001 (2002).

\bibitem{gurvin}
A. Gurevich and V.M. Vinokur, \prl {\bf 90}, ... (2003).  

\bibitem{ll}
L.D. Landau and E.M. Lifshitz, {\it Quantum Mechanics.} (Pergamon, New York, 1965). 

\bibitem{gork}
L.P. Gor'kov and T.K. Melik-Barkhudarov, Sov Phys. JETP {\bf 18}, 1031 (1964).

\bibitem{ryzik}
I.S. Gradshtein and I.M. Ryzhik, {it Tables of Integrals, Series, and Products}, (Academic Press, New York, 1980).


\end{references}
\end{document}